\documentclass[twocolumn,apj]{openjournal}

\usepackage{graphicx}

\usepackage[scaled=1.005]{newtxtext}
\usepackage{newtxmath}

\usepackage{amssymb}
\usepackage{xspace}

\usepackage[dvipsnames]{xcolor}
\usepackage[backref, pagebackref=false, hyperindex=true, breaklinks=true, colorlinks=true, urlcolor=magenta, linkcolor=magenta, citecolor=NavyBlue, citebordercolor={0 1 0}, pagecolor=red, bookmarks=true, filecolor=blue, bookmarksopen=true, plainpages=false, pdfpagemode=UseThumbs]{hyperref}

\usepackage{orcidlink}

\begin{document}

\title{From spirals to rings: dust dynamics in gravitoturbulent protoplanetary discs after late infall}

\shorttitle{Dust rings in gravitoturbulent discs after late infall}

\shortauthors{Baruteau et al.}

\author{Clément Baruteau$^1$\orcidlink{0000-0002-2672-3456}}
\author{Steven Rendon Restrepo$^2$\orcidlink{0000-0002-4299-7211}}
\author{Gaylor Wafflard-Fernandez$^3$\orcidlink{0000-0002-3468-9577}}
\author{Sebastián Pérez$^4$\orcidlink{0000-0003-2953-755X}}
\author{Zhaohuan Zhu$^5$\orcidlink{0000-0003-3616-6822}}
\author{\\Hans Baehr$^6$\orcidlink{0000-0002-0880-8296}}
\author{Farzana Meru$^7$\orcidlink{0000-0002-3984-9496}}
\author{Jean-François Gonzalez$^8$\orcidlink{0000-0001-9423-6062}}
\author{Sergei Nayakshin$^9$\orcidlink{0000-0002-6166-2206}}
\author{Sahl Rowther$^7$\orcidlink{0000-0003-4249-4478}}

\affiliation{$^1$ IRAP, Université de Toulouse, CNRS, CNES, Toulouse, France} 
\affiliation{$^2$ Leibniz-Institut für Astrophysik Potsdam (AIP), Potsdam, Germany}
\affiliation{$^3$ IPAG, Université Grenoble Alpes, CNRS, Grenoble, France}
\affiliation{$^4$ Departamento de Física, and Millennium Nucleus on Young Exoplanets and their Moons (YEMS), Universidad de Santiago de Chile, Santiago, Chile}
\affiliation{$^5$ Department of Physics and Astronomy, University of Nevada, Las Vegas, USA}
\affiliation{$^6$ Max Planck Institut für Astronomie, Heidelberg, Germany}
\affiliation{$^7$ Department of Physics, University of Warwick, Coventry, UK}
\affiliation{$^8$ Université Claude Bernard Lyon 1, ENS de Lyon, CNRS, CRAL, Saint-Genis-Laval, France}
\affiliation{$^9$ School of Physics and Astronomy, University of Leicester, Leicester, UK}

\begin{abstract}\noindent
Substructures appear to be a common trait of all extended protoplanetary discs. They are found notably in discs still surrounded by ambient, possibly infalling material. In this study, we revisit the observational signatures of gravitational instability (GI) in the dust and gas emission of protoplanetary discs. We focus on discs undergoing a short-lived episode of late infall that triggers spirals via GI, in order to investigate the long-term dynamics of millimetre-sized dust. We performed 2D hydrodynamical simulations modelling gas and millimetre-sized dust in a self-gravitating disc, with a simplified treatment of stellar irradiation and infall. Results were post-processed by 3D radiative transfer calculations. During infall, GI sets in and the disc develops a gravitoturbulent state characterised by the continuous formation of spirals. Shortly after infall ends, the millimetre-sized dust remains well coupled to the gas, and the dust emission exhibits large-scale spirals in both near-infrared scattered light and continuum emission. Even long after infall has ceased, two-armed spirals are persistently excited by GI in specific regions of the disc. Their dissipation entails the formation of multiple long-lasting pressure maxima, which can be viewed as vestiges of shocks induced by the spirals. They form persistent dust rings that are more or less axisymmetrical. Consequently, once late infall ends, a massive disc can gradually evolve from a disc with spirals in the continuum into one featuring multiple, potentially lopsided bright rings in the continuum. In near-infrared scattered light, the disc initially displays multiple spirals during infall or shortly after it ceases, before ultimately exhibiting multiple rings. The persistent excitation of spirals implies, however, that the residual line-of-sight velocity, derived from line emission, shows large-scale spirals across the disc.
\end{abstract}

  
\maketitle

% =========================
\section{Introduction}
% =========================
\noindent Protoplanetary discs exhibit a fascinating variety of sizes and morphologies that is tantalisingly reminiscent of the diversity of exoplanets. Although discs are typically small ($\lesssim$ 30 au) and appear featureless, the majority (if not virtually all) of the larger ones show substructures when observed at a sufficiently high angular resolution \citep[e.g.,][]{Bae2023,Bosschaart2026}. Demographic studies based on the dust (sub)-mm and near-IR emission show how the occurrence rate and type of substructures depend on the disc (size, mass), the star (age, mass), and the presence of ambient, possibly infalling material \citep{Bosschaart2026, Bhowmik2026, Curone2026, Garufi2026}. Two characteristics of spatially-resolved discs are of particular interest for the present study: (i) there are far more discs with rings than spirals in the dust (sub)-mm emission (by approximately a factor 10, see \citealp{Bae2023}); and (ii) discs with surrounding (possibly infalling) material usually show spirals in near-IR scattered light, but no rings \citep{Garufi2026}.

Infall can be seen as a two-stage process \citep[see, e.g., the review by][]{Kuffmeier2024}. It begins with the collapse of a pre-stellar core during the first few tens of kyr, which broadly corresponds to the Class 0 / Class I stages of protostellar evolution. Cores do not necessarily collapse in isolation: the molecular cloud environment can impact the early size and shape of discs just as much as the internal properties of the core (like its magnetic field or the ionisation rate). Post-collapse, so-called late infall can further take place during the Class I / Class II stages of protostellar evolution. Tails of dust and/or gas linking discs to their surroundings, a misalignment between the inner and outer parts of discs, provide a compelling set of observational evidence for late infall, as is the case with the SU Aur Class II disc \citep{Ginski2021}. Several numerical studies have examined the impact of late infall on disc dynamics. Various approaches have been used to model late infall, which can be grouped in two main categories: axisymmetrical and non-axisymmetric infall. 

Examples of works falling into the first category (axisymmetric infall) include the 2D grid-based simulations of \citet{Zhu2012GI}, \citet{Bae2015} and \citet{Kuznetsova2022} in which infall is modelled by additional source terms in the hydrodynamic equations, as well as the 3D SPH simulations of \citet{Longarini2025}. In some studies, the incoming gas exerts a torque on the disc where it is deposited, whilst in others it exerts no torque on the disc. When they do not fragment, gravitationally unstable discs respond to infall by exciting spiral waves \citep{Zhu2012GI,Longarini2025}, whilst when self-gravity is negligible, rings form at the edges of the infalling region, which may be subject to the Rossby-Wave Instability \citep{Bae2015,Kuznetsova2022}. 

Examples of works in the second category (non-axisymmetric infall) include the 3D SPH simulations of \citet{Calcino2025}, the 2D and 3D grid-based simulations of \citet{Lesur2015} and \citet{Hennebelle2017} in which streamers form by an appropriate external boundary condition, as well as 3D grid-based simulations of the encounter between a pre-existing disc and a cloudlet of gas left from the molecular cloud \citep{Dullemond2019,Kuffmeier2021,Huhn2026}. These works all have in common the formation of multiple spirals. A misalignment between the disc parts located inside and outside the infall region can also form \citep{Kuffmeier2021}, which is reminiscent of the simulation of the collapse of a giant molecular cloud presented in \citet{Bate2018}.

The present study builds on the work of \citet{Zhu2012GI} by investigating the long-term evolution of a self-gravitating disc after a last episode of late infall. The physical model and numerical setup of our hydrodynamical simulations are described in \S\ref{sec:model}. The main result of this study, which is that millimetre-sized dust first develops spirals -- just like the gas -- but then gradually forms multiple rings, is presented in \S\ref{sec:resultshydro}. Our simulations results were post-processed by radiative transfer calculations to produce synthetic images of the continuum intensity and of the C$^{18}$O (2-1) line emission at 1.3 mm wavelength, and of the polarised intensity at 1.65 $\mu$m. The methodology and results of the radiative transfer calculations are described in \S\ref{sec:RT}. Finally, a discussion of the results and concluding remarks follow in \S\ref{sec:conclusion}.

% =========================
\section{Model}
\label{sec:model}
% =========================
\noindent This study examines young, still massive protoplanetary discs subject to infall-driven\footnote{In the course of this work, we have tested various disc models to initiate GI through cooling rather than infall: discs with radiative cooling and an effective optical depth linked to Rosseland opacities, discs with cooling time proportional to the orbital period such that GI would be triggered from the inside out (e.g., \citealp{Baruteau2011a}). We have also tested discs with truncated inner and outer edges. In the end, we have found it more physically motivated, and more practical from a numerical point of view with less sensitivity to resolution and boundary effects, to trigger GI by controlled, localised mass infall onto the disc. In our model, infall only applies to gas, and not to dust.} gravitational instability (GI). We consider discs that do not fragment but develop a state of gravitoturbulence driven by the formation of spirals in the gas. Our aim is to investigate how dust responds to the gravitoturbulence, focusing on millimetre-sized dust particles that are marginally coupled to the gas via gas drag. For this, we have carried out two-dimensional (2D) hydrodynamical simulations to model the evolution of gas and dust in a self-gravitating disc. Simulations were performed with the public code \href{https://github.com/charango/dustyfargoadsg}{Dusty FARGO-ADSG}, an extended version of the original 2D grid-based code FARGO \citep{Masset2000} that features optional gas self-gravity, energy equation and dust dynamics. Simulations use cylindrical polar coordinates $\{R,\varphi\}$ centred on the star, with $R \in [0.2-5.0]\,R_0$ and $\varphi \in [0-2\pi]$. The reference radius $R_0$ is arbitrarily taken to be 100 astronomical units (au). We denote by $T_0$ the orbital period at $R_0$. In the following, whenever time is expressed in orbits it will implicitly correspond to orbits at $R_0$.

% -------------------------------------------
\subsection{Mass infall}
\label{subsec:infall}
% -------------------------------------------
\noindent To model mass infall in our simulations, we have adopted the same strategy as that used in the 2D self-gravitating discs simulations by \citet{Zhu2012GI}, and which we have also recently used in \citet{Nayakshin2026} and \citet{LuyaoZhang2026}. The accreted mass falls axi-symmetrically on a rather narrow range of orbital radii in the disc (between $R_{\rm a}$ and $R_{\rm b}$), with a uniform mass accretion rate $\dot{M}_{\rm in}$. Following \citet{Zhu2012GI}, the infalling mass is assumed to have the same specific angular momentum as the background gas where it flows in, such that mass infall increases the mass, the internal energy of the disc, but it does not torque it. Mass infall thus implies a source term $\dot{\Sigma}_{\rm in}$ in the continuity equation for the gas surface density $\Sigma$,  which is given by equation 7 of \citet{Zhu2012GI}:
\begin{equation}
\dot{\Sigma}_{\rm in}(R) = \frac{\dot{M}_{\rm in}}{2\pi R^2} \times (2-\sigma) \times \frac{R^{2-\sigma}}{R_{\rm b}^{2-\sigma} - R_{\rm a}^{2-\sigma}},
\label{eq:sigmadot_in}
\end{equation}
where $\sigma = 1$ is the opposite of the power-law exponent used in the expression for the initial gas surface density profile (see \S\ref{subsec:initialconditions}). Mass infall also implies a heating source term $\dot{e}_{\rm in}$ in the equation for the gas thermal energy density $e$, given by equation 8 of \citet{Zhu2012GI}:
\begin{equation}
\dot{e}_{\rm in}(R) = \frac{k_{\rm B}\dot{\Sigma}_{\rm in}(R)T_{\rm irr}(R)}{(\gamma-1)\mu m_{\rm p}},
\label{eq:edot_in}
\end{equation}
with $k_{\rm B}$ the Boltzmann constant, $T_{\rm irr}$ the temperature profile set by stellar irradiation, which we fix to $T_{\rm irr} (R) = 21.5\,{\rm K} \times (R/R_0)^{-1/2}$, $\gamma = 11/7$ is the 2D adiabatic index \citep{Zhu2012GI}, $\mu=2.4$ is the mean molecular weight, and $m_{\rm p}$ the proton's mass. In the simulation presented in \S\ref{sec:resultshydro}, $R_{\rm a} = 0.85 R_0$, $R_{\rm b} = R_0$ and $\dot{M}_{\rm in} = 10^{-5} M_{\odot}\,{\rm yr}^{-1}$. In practice, $\dot{\Sigma}_{\rm in}$ smoothly increases from the start of the simulation over $T_0$, then gradually returns to zero after 30 $T_0$. We thus model a last, finite episode of late infall onto the disc. As will be shown in \S\ref{sec:resultshydro}, it increases the disc mass by approximately 0.3 $M_{\odot}$.

% -----------------------------
\subsection{Self-gravity}
% -----------------------------
\noindent Our simulations include gas self-gravity \citep{BaruteauMasset2008b} but discard dust self-gravity (because the dust-to-gas mass ratio remains small in our simulations; see \S\ref{subsec:spacetime}). The components of the self-gravitating acceleration are calculated via fast Fourier transforms using the Bessel kernel\footnote{In 2D simulations where GI is triggered by cooling, the Bessel kernel allows to reach a well-defined numerical convergence for the critical cooling time below which fragmentation occurs \citep{RendonRestrepo2026b}.} recently derived from first principles in \citet{RendonRestrepo2026}, instead of the original kernel based on a Plummer formulation of the self-gravitating potential with a smoothing length. Since our simulations use a reference frame in which the star is fixed, the disc also feels an indirect acceleration which is the opposite of the gravitational acceleration exerted by the disc on the star (the so-called disc's indirect term; see for instance \citealp{Crida2025a}). Our study assumes that the disc evolution is solely driven by radial turbulent transport due to gravitoturbulence. There is therefore no additional explicit viscosity nor thermal diffusivity in our simulations.

% -----------------------------
\subsection{Energy equation}
\label{subsec:eeq}
% -----------------------------
\noindent The equation for the thermal energy density $e$ reads
\begin{equation}
\frac{\partial e}{\partial t} + \nabla\cdot(e {\bf v}) + (\gamma-1)e\nabla\cdot {\bf v} = Q^{+}_{\rm bulk} - \frac{2\sigma_{\rm SB} (T^4 - T^4_{\rm irr})}{\tau_{\rm eff}} + \dot{e}_{\rm in}.
\label{eq:energy_eqn}
\end{equation}
The left-hand side of Eq.~(\ref{eq:energy_eqn}) encapsulates the advection of specific entropy, with ${\bf v} = (v_R, v_{\varphi})^{\rm T}$ the gas velocity. On the right-hand side of the equation, the first term ($Q^{+}_{\rm bulk}$) represents artificial heating\footnote{Regarding the GI spirals in our simulations, we find that $Q^{+}_{\rm bulk}$ is generally lower than the compressional heating term (the last term on the left-hand side of Eq.~\ref{eq:energy_eqn}), but not by a significant margin.} due to shocks via the use of a Von Neumann-Richtmyer artificial bulk viscosity \citep{StoneNorman1992a}. The second term models both radiative cooling through the disc surfaces and stellar irradiation heating. It can be seen as a relaxation of the disc temperature $T$ towards the temperature profile $T_{\rm irr}(R)$ set by stellar irradiation (in the absence of gravitoturbulence and infall, $T(R) = T_{\rm irr}(R)$ in a steady state). In this term, $\sigma_{\rm SB}$ is the Stefan-Boltzmann constant, and $\tau_{\rm eff}$ is an effective optical depth that is a linear combination of the optical depth $\tau = \kappa_{\rm R} \Sigma/2$ and its inverse. The expression for $\tau_{\rm eff}$ can be found in equation~4 of \citet{Marzari2012}. The Rosseland mean opacities $\kappa_{\rm R}(\Sigma,T)$ are computed from \citet{BellLin1994}. Finally, the last term in Eq.~(\ref{eq:energy_eqn}) is the heating rate due to infall, given by Eq.~(\ref{eq:edot_in}). 

% -----------------------------
\subsection{Dust}
% -----------------------------
\noindent In Dusty FARGO-ADSG, dust can be modelled as a single dust fluid of a given size \citep{Zhu2012DustFluid}, and/or as Lagrangian super-particles with a size distribution \citep{Baruteau2016, Baruteau2019}. We have tested both approaches, which have their pros and cons in terms of resolution, computational efficiency, and potential sensitivity to boundary conditions. These two approaches yield consistent results, except that the dust density field reconstructed from the super-particles position naturally contains more noise than that of a dust fluid, which affects synthetic images of the dust emission. In the end, we only present results with a zero-pressure dust fluid modelling solid particles of 1 mm in size, a specific size used to produce synthetic images of the continuum emission (see \S\ref{subsec:RTsetup}). 

Dust is dragged by the gas (Epstein regime) but the dust drag on the gas is discarded for simplicity and because of small dust-to-gas density ratios in our simulations. For the initial radial profiles of the gas temperature and surface density in our simulations (see \S\ref{subsec:initialconditions}), the mean free path $\lambda_{\rm mfp}$ of the gas molecules is $\approx$ 40 m at $R = 0.2 R_0 = 20$ au, where it is at its minimum (for the geometrical cross-section, we assume it to be $\approx 7 \times 10^{-20}$ m$^2$, as in \citealp{Rice2004}). The mean free path then increases rapidly with $R$ ($\lambda_{\rm mfp} \propto R^{9/4}$ up to $R_0$). Therefore, it is much greater than the size of the solids considered in this study, justifying the use of the Epstein regime to model the drag force. Furthermore, to save computing time, the short-friction time approximation is used since pretty small Stokes numbers are reached in the disc's inner parts (around $10^{-3}$). Consequently, dust feels the gas self-gravity via gas drag. When computing the Stokes number, an internal mass volume density of 1.3 g cm$^{-3}$ is assumed for the dust.

% -----------------------------
\subsection{Initial conditions}
\label{subsec:initialconditions}
% -----------------------------
\noindent {\it Gas -- } The initial radial profile of the disc temperature is chosen to match $T_{\rm irr} (R)$. In simulations carried out with the FARGO family of codes, the initial temperature is set by the initial aspect ratio $h_0$, that is the ratio between the (initial) sound speed and the Keplerian velocity. Our $T_{\rm irr} (R)$ profile (see \S\ref{subsec:infall}) corresponds to $h_0(R) = 0.092 \times (R / R_0)^{1/4}$, given that the central star has a solar mass. The initial radial profile of the gas surface density takes the form $\Sigma_0(R) = \Sigma_0(R_0) \times (R/R_0)^{-\sigma} f(R)$ with $f(R) = 1$ if $R \leq R_{\rm b}$, and $f(R) = \exp(1 - R/R_{\rm b})$ if $R>R_{\rm b}$. The $f$ function thus mimics the effect of an exponential cutoff for $R \geq R_{\rm b}$. As anticipated in \S\ref{subsec:infall}, we take $\sigma=1$. At our reference radius $R_0=100$ au, the initial surface density is $\Sigma_0(R_0) \approx 13$ g cm$^{-2}$ (or 0.0148 in code units, where $R_0$ is the length unit and $M_{\odot}$ the mass unit). The initial disc-to-star mass ratio is about 0.16, and the minimum Toomre-Q parameter is $\approx$ 2.7 at $R=R_0$. A small non-axisymmetric perturbation is added to the initial surface density of the gas, at the 0.1\% relative level. In the simulation presented in \S\ref{sec:resultshydro}, this perturbation is a mode with azimuthal wavenumber $m=2$. It implies that the GI mode that grows fastest also has $m=2$. We have verified that the use of white noise can lead to a different fastest growing mode, but either way the non-linear saturation of GI results in discs with very similar morphology, featuring prominent $m=2$ spirals. The initial azimuthal velocity of the gas is calculated such that the disc be in centrifugal balance. The initial radial velocity is null.
\\
\par\noindent {\it Dust -- } The dust density profile is initialised such that the dust-to-gas density ratio equals 1 percent. The initial dust velocity is Keplerian.

% -----------------------------
\subsection{Numerics and boundary conditions}
\label{subsec:numericsandbc}
% -----------------------------
\noindent The computational grid has 840 cells in the radial direction and 1600 cells in the azimuthal direction. A logarithmic radial spacing is used, and for our chosen resolution cells are approximately square across the grid. There are about 25 cells per pressure scale-height in each direction. Note that a logarithmic radial spacing is required for the components of the self-gravitating acceleration to read as convolution products and be computed with fast Fourier transforms \citep{BaruteauMasset2008b}. The Bessel kernel used to compute the self-gravitating acceleration features the gas pressure scale height \citep{RendonRestrepo2026}, which evolves in time as does the gas temperature. However, our disc model and our choice of Rosseland opacities imply that the background temperature evolves relatively mildly over time (see the bottom-left panels in Figures~\ref{fig:fig_spacetime2} and~\ref{fig:fig_extra}): the relative variation in the background temperature profile does not exceed 10\%, and that of the pressure scale-height therefore does not exceed 5\%. To save computing time, the Bessel kernel is thus computed only once at the beginning of the simulation: it is not updated at every hydrodynamical timestep. Given the small variations in the pressure scale-height in our simulation, retaining the initial Bessel kernel should have little impact on the calculation of the self-gravitating acceleration. Still, future work should generally assess the expected accuracy of this approximation. The simulation presented in \S\ref{sec:resultshydro} was run over 500 orbits. It took about 5500 CPU hours to complete.

Boundary conditions have turned out to be sensitive, perhaps not surprisingly. On the one hand, outflow boundaries allow gas and dust to flow out of the grid, but cause a certain degree of wave reflection at its edges. They sometimes lead to the formation of clumpy structures at the edges of the grid, which can significantly reduce the hydrodynamical timestep when they form at the inner edge. On the other hand, the simultaneous use of damping layers located near the edges of the grid (where the fields are damped toward their initial radial profile) reduce or even prevent wave reflections, but they also prevent the disc from flowing out of the grid, so that the mass of the disc does not decrease (or decreases only slightly) over time. Also, excessive damping might favour the formation of standing modes in the grid which, along with the large density of the gas at the inner edge, may be prone to the reflex instability \citep{Crida2025b, Magnan2026b}. We witnessed the reflex instability in several preliminary simulations using damping layers, which caused the disc to become massively eccentric. In the end, for the simulation in \S\ref{sec:resultshydro}, we decided to use outflow boundaries without damping layers. For the gas, we found good results by using azimuthally-averaged quantities at both the inner and outer edges. This means that in each ghost ring, the density and the thermal energy density are set equal to their azimuthally-averaged counterpart in the adjacent active ring. We do the same for the radial velocity, but only if the azimuthally-averaged radial velocity in the active ring is negative; otherwise the radial velocity in the ghost ring is set to zero. Lastly, the azimuthal velocity in the ghost rings is a Keplerian extrapolation of the azimuthally-averaged azimuthal velocity in the adjacent active rings. For the dust fluid, we use the same boundary conditions as above for the dust's density and velocity in the inner ghost ring. The dust fields in the outer ghost ring retain their initial (axisymmetric) value.

Finally, the mm dust fluid initially drifts inwards quite rapidly in the outer parts of the disc, where the Stokes number approaches unity. This can cause the dust density to take extremely small values which the code does not handle well. To avoid this, we use floor values for the dust density, which we set to $10^{-5}$ times the local initial dust density. This means that, in each grid cell, the instantaneous-to-initial dust density ratio cannot fall below $10^{-5}$. This very small ratio ensures that there is no significant artificial increase in the dust mass over time.

% =========================
\section{Results of hydrodynamical simulation}
\label{sec:resultshydro}
% =========================
\noindent We present in this section the results of one simulation, for which mass infall is set over 30 $T_0$ between 0.85 $R_0$ and $R_0$, i.e. between 85 and 100 au, with an accretion rate of $10^{-5} M_{\odot}\,{\rm yr}^{-1}$. The infall therefore lasts for about $3\times10^4$ yr (0.03 Myr). These parameters are found to produce a gravitoturbulent disc with no fragmentation, in agreement with the simulation labelled R100$\_$1e-5 in \citet{Zhu2012GI}, for which $\dot{M}_{\rm in}$, $R_{\rm a}$, $R_{\rm b}$, $R_0$ and $\Sigma_0(R_0)$ take the same values as in our model. 

In order to better understand why no fragmentation is obtained, we propose to calculate a dimensionless timescale for the increase in the disc's surface density due to mass infall, in the same way as the "beta" parameter $\beta_{\rm cool}$ associated with the cooling timescale in discs where GI is regulated by cooling: $\beta_{\rm in}(R) = \Omega(R) \times \Sigma(R) / \dot{\Sigma}_{\rm in}(R)$. Since in our model mass infall occurs near $R_0$, we compute $\beta_{\rm in}$ at $R_0$. Using Eq.~(\ref{eq:sigmadot_in}) and the parameters of our model, we obtain $\beta_{\rm in} (R_0) \approx 8.6$. {\it If} the criterion for the onset of fragmentation was the same for infall- and cooling-driven GI, namely $\beta_{\rm in} \lesssim 3-4$ \citep[e.g.,][]{Gammie01,Paardekooper12}, then we should not expect fragmentation in our model. Interestingly, by increasing $\dot{M}_{\rm in}$ by a factor of 3, thereby implying $\beta_{\rm in} (R_0) \approx 2.9$, we do find fragmentation in the gas ring where infall is ongoing. More specifically, we find compact clumps that form near $R_{\rm a}$ and then zip through the inner edge of the computational domain in less than 6 $T_0$. It is consistent with \citet{Zhu2012GI}, who found marginal clump formation for $\dot{M}_{\rm in} = 3\times 10^{-5} M_{\odot}\,{\rm yr}^{-1}$ in their simulation labelled R100$\_$3e-5. Although this remains to be consolidated in future work, our results indicate that infall-driven GI would trigger fragmentation when $\beta_{\rm in} \lesssim 3-4$ where infall takes place. Whatever the exact critical value of $\beta_{\rm in}$ for fragmentation to occur via infall, a criterion like above means that, when the mass infall rate is too high, the disc does not have enough time to generate waves (gravitoturbulence) capable of efficiently evacuating this material, and it ends up fragmenting. Some more insight is provided in \S\ref{sec:beta_infall} of the Appendix.

% -------------------------------------------
\subsection{Overall evolution}
\label{subsec:overall}
\begin{figure}
    \centering
        \includegraphics[width=0.99\hsize]{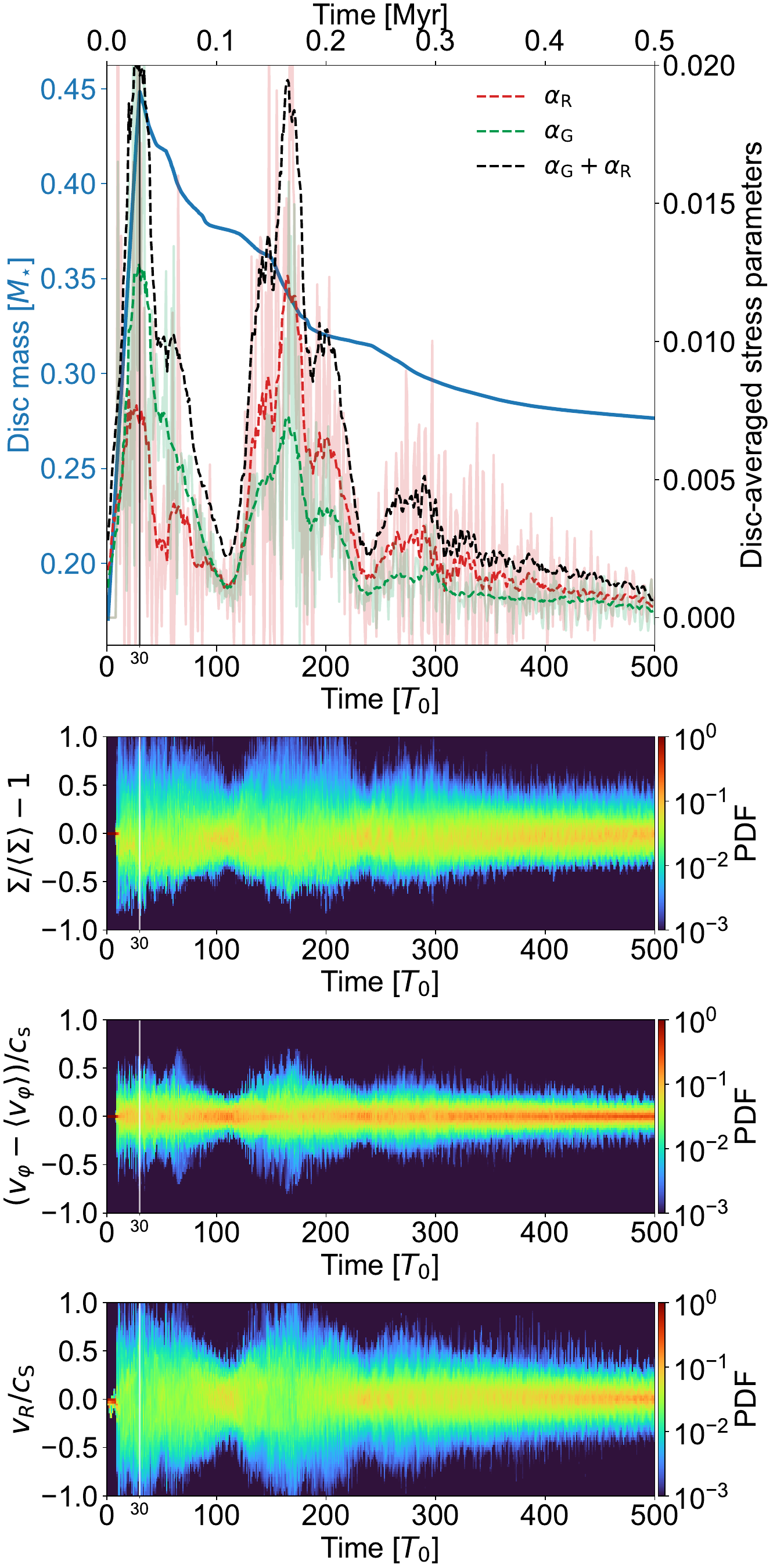}
    \caption{Top panel: time evolution of the mass of the disc gas (solid blue curve; left y-axis), and of the disc-averaged Reynolds and gravitational stresses (red and green transparent solid curves, respectively; right y-axis). The dashed curves show the disc-averaged stresses smoothed over 10 $T_0$ (the dashed black curve shows the smoothed total stress). Second to fourth panels from the top: time evolution of the normalised probability density function (PDF) of the non-axisymmetric perturbation of the gas density relative to its azimuthal average, of the non-axisymmetric perturbation of the gas azimuthal velocity relative to the sound speed, and of the gas radial velocity relative to the sound speed. In all the panels, the vertical line indicates the moment when the infall stops.
    } 
    \label{fig:fig_discmass}
\end{figure}

% -------------------------------------------
In the upper panel of Figure~\ref{fig:fig_discmass}, the solid blue curve shows the time evolution of the disc-to-star mass ratio (only the mass of the gas is accounted for). Due to infall, the disc-to-star mass ratio increases, here linearly, from $\sim$0.16 to $\sim$0.45. About 0.03 Myr after the beginning of the simulation, infall stalls and the disc mass starts decreasing over time as gas is allowed to flow out of the grid (see \S\ref{subsec:numericsandbc}). The disc-to-star mass ratio decreases continuously, though non monotonically, and reaches $\sim$0.28 at the end of the simulation at 0.5 Myr. Most of the disc mass leaves the computational grid through its inner edge. Whether this material would be ultimately accreted onto the star, or evacuated by winds, is not modelled in this work. Some of this material could, in fact, contribute to an increase in stellar mass, which would lead to a more rapid decrease in the disc-to-star mass ratio than in our simulation, thereby shortening the GI phase.

%FFFFFFFFFFFF
%FFFFFFFFFFFF
The transparent solid curves further display the disc-averaged alpha parameters associated with the Reynolds and gravitational stresses ($\alpha_{\rm R}$ and $\alpha_{\rm G}$). Each of them is calculated as a radial average of its radial profile. The radial profile of the Reynolds alpha parameter is given by equation A1 of \citet{Baruteau2011a}:
\begin{equation}
\alpha_{\rm R}(R) = \frac{2}{3} \frac{\langle \Sigma \delta v_R \delta v_{\varphi} \rangle}{\langle \Sigma c_{\rm s}^2 \rangle},
\label{eq:alphareynolds}
\end{equation}
where $\langle \cdot \rangle$ denotes the azimuthal average, $\delta v_R = v_R - \langle v_R \rangle$, $\delta v_{\varphi} = v_{\varphi} - \langle v_{\varphi} \rangle$, and $c_{\rm s}$ is the sound speed. The radial profile of the gravitational alpha parameter is given by equation A2 of \citet{Baruteau2011a}:
\begin{equation}
\alpha_{\rm G}(R) = \frac{2}{3} \frac{\langle  \int_{-\infty}^{\infty} (4\pi G)^{-1} g_R\,g_{\varphi}\,dz \rangle}{\langle \Sigma c_{\rm s}^2 \rangle},
\label{eq:alphagravitational}
\end{equation}
with $G$ the gravitational constant, and where $g_R$ and $g_{\varphi}$ are respectively the radial and azimuthal self-gravitating accelerations. For the dependence of $g_R$ and $g_{\varphi}$ on the vertical coordinate $z$, the reader is referred to appendix A in \citet{Baruteau2011a}. Only in Eq.~(\ref{eq:alphagravitational}) does the calculation of $g_R$ and $g_{\varphi}$ feature a smoothing length. The latter is taken to be 0.6$H(R_0)\times R/R_0$, as it is what best reproduces the use of a Bessel kernel \citep{RendonRestrepo2026}. A 10-point Gauss-Legendre integration is used for the vertical integration. We point out that very similar values for $\alpha_{\rm G}$ were obtained by using the $g_R$ and $g_{\varphi}$ fields provided directly by the simulation, and by replacing the vertical integral in Eq.~(\ref{eq:alphagravitational}) with a multiplication by 2H($R_0$). 

Overall, we see that $\alpha_{\rm R}$ and $\alpha_{\rm G}$ take on similar\footnote{In the cooling-driven GI disc simulations of \citet{Baruteau2011a}, $\alpha_{\rm R}$ and $\alpha_{\rm G}$ also took on similar values. These simulations had initial disc-to-star mass ratios $\sim$0.4, thus very similar to that of our simulation near the onset of GI.} values and show a similar trend over time. This is better seen with the dashed curves in the panel, which are obtained by smoothing the alpha parameters over a time window of 10 $T_0$. Both smoothed quantities first increase rapidly and peak when the disc mass has reached its maximum value. They then decrease as the disc mass decreases. Interestingly, from about 100 $T_0$, an episode of reinvigorated GI sets in which causes a new increase in $\alpha_{\rm R}$ and $\alpha_{\rm G}$, as well as a faster decrease in the disc mass. We note that from this episode onward, $\alpha_{\rm R}$ becomes larger than $\alpha_{\rm G}$. A final, less prominent upturn in GI activity is visible around 280 $T_0$, after which both alpha parameters decrease rather smoothly over time. During the last 200 orbits, the quantity $\alpha_{\rm R} + \alpha_{\rm G}$ averages out to about $2\times 10^{-3}$. 

The time variability of the disc-averaged stresses is further illustrated with the three lower panels in Figure~\ref{fig:fig_discmass}. The second panel from the top displays the probability density function (PDF) of the quantity $\Sigma / \langle\Sigma\rangle - 1$, which corresponds to the non-axisymmetric perturbation of the gas surface density relative to its azimuthal average (we also denote it by $\delta\Sigma / \langle\Sigma\rangle$; note that all PDFs are normalised relative to their maximum value). When that quantity cancels out, $g_{\varphi}$ cancels out and so does $\alpha_{\rm G}$. We see that, not surprisingly, larger values of $\alpha_{\rm G}$ are obtained when the PDF of $\delta\Sigma / \langle\Sigma\rangle$ has broader width and/or takes maximum values for non-vanishing $\delta\Sigma / \langle\Sigma\rangle$. This is especially visible during the three aforementioned episodes of more intense GI activity (around $30 T_0$, $150 T_0$ and $280 T_0$). Similarly, the third panel shows the normalised PDF of the quantity $\delta v_{\varphi} / c_{\rm s}$, which corresponds to the non-axisymmetric perturbation of the gas azimuthal velocity relative to the sound speed. We see that the PDF of $\delta v_{\varphi} / c_{\rm s}$ remains centred around 0, contrary to that of $\delta\Sigma / \langle\Sigma\rangle$, and that larger values of $\alpha_{\rm R}$ are obtained when the PDF of $\delta v_{\varphi} / c_{\rm s}$ has larger width. Finally, the fourth panel displays the normalised PDF of the gas radial velocity relative to the sound speed. It can be seen that the PDF of $v_R / c_{\rm s}$ has a width roughly twice that of the PDF of $\delta v_{\varphi} / c_{\rm s}$. Interestingly, \citet{Longarini2021} showed that, in a gravitoturbulent disc, $v_R / \delta v_{\varphi} \approx 4 m M_{\rm disc}/M_{\star}$, where $m$ is the azimuthal wavenumber of the dominant mode and $M_{\rm disc}$ is the mass of the disc. In our simulation, the dominant mode has $m=2$ (see also Figure~\ref{fig:fig_mode}), and, when this is the case, the disc-to-star mass ratio varies between approximately 0.3 and 0.45. As a result, $v_R / \delta v_{\varphi}$ should vary approximately between 2.5 and 3.5, which is consistent with the relative widths of the PDFs of $v_R$ and $\delta v_{\varphi}$.

% -------------------------------------------
\subsection{Space-time diagrams}
\label{subsec:spacetime}
% -------------------------------------------
%FFFFFFFFFFFF
\begin{figure*}
    \centering
        \includegraphics[width=\hsize]{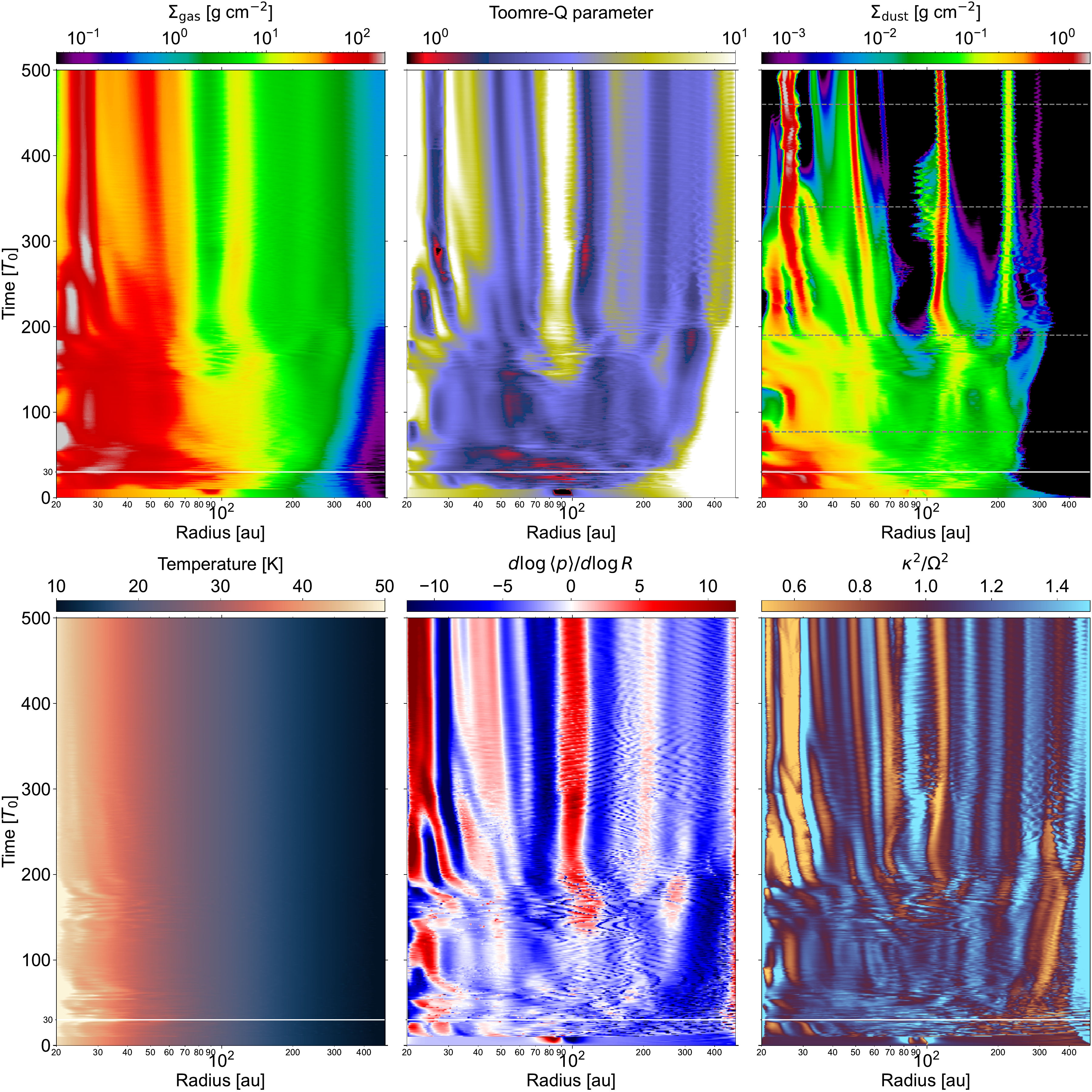}
    \caption{Space-time diagrams of azimuthally-averaged radial profiles of several disc quantities. From top-left to bottom-right are displayed the gas surface density ($\Sigma_{\rm gas}$), the Toomre-Q parameter (see text), the dust surface density ($\Sigma_{\rm dust}$), the temperature, the logarithmic radial gradient of the gas thermal pressure ($d\log \langle p\rangle/d\log R$), and the squared ratio between the gas epicyclic frequency $\kappa$ and the angular frequency $\Omega$ (ratio that we more simply refer to as the gas normalised vorticity). In each panel, time is shown on the y-axis with a linear scale (the sampling timescale is one orbit), while orbital radius is shown on the x-axis with a logarithmic scale. In the top-right panel, horizontal dashed lines mark the times corresponding to the panels shown in Figures~\ref{fig:fig_2Ddens} and~\ref{fig:fig_RTmedley}-\ref{fig:fig_mom1maps}. In all the panels, the horizontal solid line indicates the moment when the infall stops.
    } 
    \label{fig:fig_spacetime2}
\end{figure*}
%FFFFFFFFFFFF
%FFFFFFFFFFFF
\begin{figure*}
    \centering
        \includegraphics[width=\hsize]{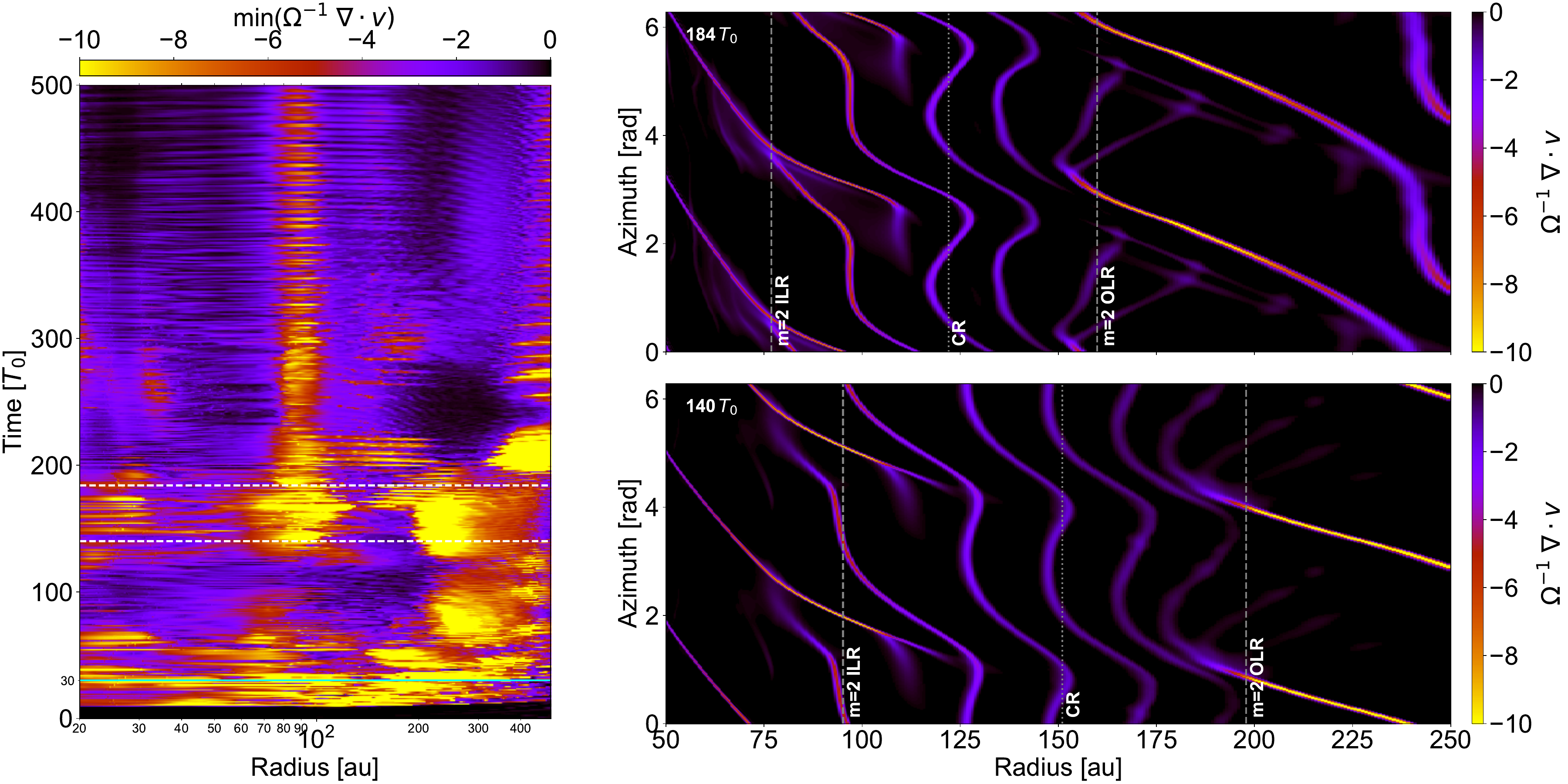}
    \caption{Left: space-time diagram of the minimum, at each orbital radius, of the velocity divergence divided by the angular frequency. This quantity indicates where spirals (shocks) are excited and may dissipate (see text). The horizontal solid line indicates the moment when the infall stops. Right: ratio between the velocity divergence and the angular frequency in polar coordinates, at the two times marked by horizontal dashed lines in the left panel. They show the excitation of a two-armed spiral, whose corotation radius (CR), the inner and outer Lindblad resonances (ILR, OLR) are respectively shown by dotted and dashed lines.
    } 
    \label{fig:fig_spacetime2b}
\end{figure*}
%FFFFFFFFFFFF
\noindent We continue our description of the overall behaviour of the disc in our simulation by showing in Figure~\ref{fig:fig_spacetime2} space-time diagrams of azimuthally-averaged radial profiles for several disc quantities. The radial profile of the gas surface density is displayed in the top-left panel. Apart from the infall region, where it increases continuously, the gas density remains relatively smooth and largely unperturbed during the first 9 to 10 orbits, after which the gas density profile changes abruptly. This change is due to the onset of GI, which triggers the formation of large-scale spirals. The dominant azimuthal wavenumber of the spirals is $m=2$, in agreement with the 3D infall-driven GI simulations of \citet{Longarini2025} for the range of disc-to-star mass ratios covered in our simulation. The spirals cause a sudden and substantial transport of mass and angular momentum through the disc, which is manifested in the large initial increase in the alpha parameters seen in Figure~\ref{fig:fig_discmass}. This transport alters the background density profile, which nevertheless remains relatively smooth and monotonous up to about 150 $T_0$. From this point onwards, which coincides with the peak in the reinvigorated GI episode seen in Figure~\ref{fig:fig_discmass}, the gas density gradually forms several gaps and bumps: a more pronounced gap begins to form around 90 au, followed by other gaps around 35 and 180 au. These bumps and gaps, which can also be seen in the top-left panel of Figure~\ref{fig:fig_extra} in \S\ref{sec:extrafigures} of the Appendix, are due to the continuous excitation and dissipation of spirals.

To show this, we proceed directly to Figure~\ref{fig:fig_spacetime2b}. The left panel displays the space-time diagram of the minimum, at each orbital radius, of the velocity divergence divided by the angular frequency. (It means that at each orbital radius, at a given time, we take the minimum along the azimuthal direction of the ratio between the gas' velocity divergence and angular frequency; we now refer to this ratio as the normalised velocity divergence). The reason for looking at this quantity is that it indicates where the spirals are excited and may deposit the fluxes of energy and angular momentum that they carry \citep{Longarini2025}. By comparing the space-time diagram in Figure~\ref{fig:fig_spacetime2b} with that of the gas density in Figure~\ref{fig:fig_spacetime2}, we can see that local minima in $\Sigma_{\rm gas}$ (i.e., gaps) coincide pretty well with strong minima in the normalised velocity divergence. This is particularly clear for the more prominent gap around 90 au. Likewise, local maxima in $\Sigma_{\rm gas}$ tend to coincide with small (or weak) minima in the normalised velocity divergence.

As far as the gaps around 90 au and 180 au are concerned, we attribute them to the continuous excitation of a two-armed spiral density wave in the disc region between 120 and 160 au. This is illustrated by the two panels on the right-hand side of Figure~\ref{fig:fig_spacetime2b}, which are inspired by Figure 12 of \citet{Longarini2025}. The panels display the normalised velocity divergence at 140 $T_0$ and 184 $T_0$ in a large fraction of the disc. To get the corotation radius of the spirals, we restarted the simulation at the two aforementioned times by producing frequent outputs (every twentieth of an orbit) and by simply measuring the angular pattern frequency of the spirals. The corotation radius of the spirals is marked by a dotted line in the panels, and the associated $m=2$ inner and outer Lindblad resonances are shown by dashed lines. In agreement with \citet{Longarini2025}, the launching point of the outer spirals coincides rather well with the outer $m=2$ Lindblad resonance, while for the inner spirals the launching point can largely approach the corotation radius. Unlike \citet{Longarini2025}, the corotation radius does not coincide with the radius or with the range of radii from where infall sets in, which is not surprising given that we are examining moments in the simulation long after the infall has ceased. Evanescent waves are clearly visible between the Lindblad resonances, with a peculiar zigzag shape. It is possible that this shape reflects the temporal variability of the spirals' corotation radius, but this would require a dedicated study. In any case, we stress that the spirals remain active all over the time interval considered here (between 140 and 184 $T_0$), and their corotation radius sweeps across a fairly extended region in the radial direction with a non-trivial temporal evolution. Although a limited time interval is considered here, we stress that the spirals are actually excited over the entire duration of the simulation. We attribute the gaps in the gas density around 90 au and 180 au to the above $m=2$ spirals cumulatively depositing their angular momentum flux as they steepen into shocks near their launching radius, in some analogy with planetary wakes \citep[e.g.,][]{Bae2017}. Furthermore, we think that the gap carved around 35 au from $\sim$230 $T_0$ is likely due to the excitation of additional $m=2$ spirals in the disc's inner parts; however, the proximity of the inner edge of the computational domain, and the resulting wave reflections, do not help clarify the situation.

Going back to Figure~\ref{fig:fig_spacetime2}, the lower-left panel shows that the temperature profile remains little perturbed overall. Shock heating is more particularly visible inward of 100 au and during the episodes of more intense GI activity, when the stress parameters take their maximum value (see Figure~\ref{fig:fig_discmass}). By comparing with the left panel of Figure~\ref{fig:fig_spacetime2b}, we see that, in the inner regions of the disc, there is a good match between when and where the temperature rises most, and when and where the normalised velocity divergence has its strongest minima (i.e., when and where shocks are most intense). In the optically thick regime, the dimensionless cooling timescale, by which we mean the quantity
\begin{equation}
\beta_{\rm cool} = \Omega \tau_{\rm cool}  \sim \frac{\Sigma \tau_{\rm eff} \Omega}{(\gamma-1) \sigma_{\rm SB} T^3},
\label{eq:beta}
\end{equation}
is displayed in the lower-middle panel in Figure~\ref{fig:fig_extra} at several times in the simulation. More specifically, azimuthally-averaged radial profiles ($\langle\beta_{\rm cool}\rangle$) are shown. We see that $\langle\beta_{\rm cool}\rangle$ stays below unity outward of 70 to 100 au, which explains why shock heating is barely visible in this region and that the temperature retains its initial radial profile. We did not observe any cooling-driven fragmentation, either before or after infall has stalled, even though the Toomre-Q parameter reaches values close to unity in regions where $\langle\beta_{\rm cool}\rangle < 1$. This is because, in our model, heating by stellar irradiation is taken into account in addition to radiative cooling, so that the classical result that disc fragmentation should occur when $\beta_{\rm cool}$ is less than a few \citep[e.g.,][]{Gammie01,Paardekooper12,Zhu2012GI} is not relevant.

Next, the upper-middle panel in Figure~\ref{fig:fig_spacetime2} displays the radial profile of the Toomre-Q parameter, ${\rm Q} = c_{\rm s}\kappa/\{\pi G\Sigma_{\rm gas}\}$, with $\kappa$ the epicyclic frequency which is calculated via the azimuthally-averaged radial profile of the gas azimuthal velocity. The colour scale is chosen to highlight (in red) values of Q close to unity. We see that Q initially decreases below 1 in the infall region until the growth of the first spirals, after which Q maintains a minimum value close to 1 throughout the simulation, predominantly near the initial infall region. From about 200 orbits, the Toomre-Q profile shows persistent radial structures with local minima and maxima that reflect the radial variations in the gas density profile. These minima are between 1 and 2, implying that the disc persistently excites GI spirals throughout the simulation.

The aforementioned radial structures are also clearly seen in the lower-middle panel of Figure~\ref{fig:fig_spacetime2}, which shows the radial profile of the logarithmic radial gradient of the gas pressure. It displays alternating negative and positive values quite early in the simulation, and which become persistent beyond 200 $T_0$. These can be regarded as tracing zonal flows, and imply the formation of quasi-steady pressure maxima (where $d\log\langle p \rangle / d\log R$ goes from positive to negative values with increasing $R$). 

The formation of these persistent pressure maxima causes the gradual formation of dust rings, as is clearly seen in the dust's surface density profile in the upper-right panel of Figure~\ref{fig:fig_spacetime2}. As noted above, we attribute the gaps in the gas density to regions where the persistent $m=2$ spirals cumulatively deposit their angular momentum flux near their launching radius (i.e., near the Lindblad resonances). In our view, the pressure maxima and the associated dust rings thus form just outside the Lindblad resonances of the two-armed spirals. Focusing on the upper-right panel, we see that the dust's initial evolution much follows that of the gas, except in the disc's outer parts where dust drifts fastest due to gas drag (radial profiles of the Stokes number are shown in the bottom-right panel of Figure~\ref{fig:fig_extra}). We point out that the inward drift due to gas drag competes with outward drift due to GI-driven spirals. This is best seen by the increase in the dust surface density outward of 200 au and beyond $\sim$30 $T_0$. The tighter coupling between dust and gas in the inner parts of the disc means that the dust drifts through the inner edge of the computational domain at a similar rate to the gas. Consequently, once infall ceases, the dust-to-gas mass ratio remains nearly constant, increasing only slightly from around 0.30\% to 0.37\% by the end of the simulation. Additionally, we have checked that the dust-to-gas density ratio never exceeds 10\% in our simulation, which justifies a posteriori our choice to discard dust drag on the gas.

%FFFFFFFFFFFF
\begin{figure*}
    \centering
        \includegraphics[width=\hsize]{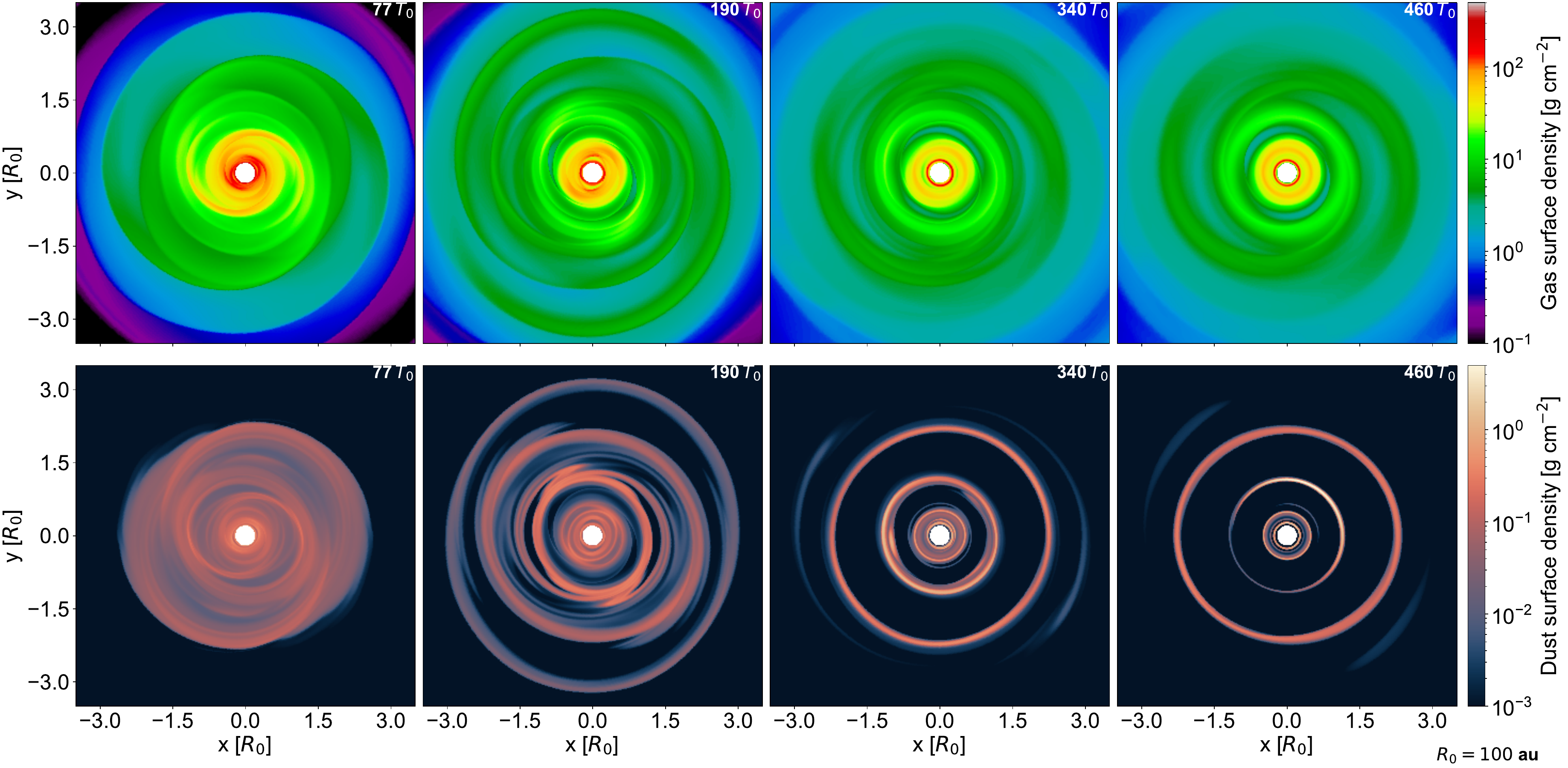}
    \caption{Surface density of the gas (top panels) and of the 1 mm-sized dust fluid (bottom panels) at four selected times, showing that the structure of the dust gradually evolves from multiple spirals to multiple rings.} 
    \label{fig:fig_2Ddens}
\end{figure*}
%FFFFFFFFFFFF
To finish with Figure~\ref{fig:fig_spacetime2}, the lower-right panel displays the radial profile of the quantity $\kappa^2/\Omega^2$, which is twice the ratio between the vorticity and the angular frequency of the gas, and which we dub {\it normalised vorticity}. The reason for showing this quantity is that vorticity can be considered as another diagnostic of shocks in discs, akin to the velocity divergence. Potential vorticity, that is the ratio between vorticity and surface density, is a more conventional diagnostic, shocks being associated with local minima in the potential vorticity profile \citep[see, e.g.,][]{WFB20}. Here we have chosen vorticity to focus on the radial structures arising from the velocity field only, and not from a combination of the gas velocity and surface density. Still, we have checked that the radial profiles of the gas vorticity and potential vorticity display the same structures; in particular, the location of their radial minima coincides well. 

The persistent minima that form in the normalised vorticity profile can be viewed as fossilised remnants of shocks due to GI spirals. They progressively build up as the mass of the disc and its level of gravitoturbulence decrease. We note again that the rapid cooling in the outer disc ($\gtrsim$100 au) implies that, although shocks are evident from the normalised vorticity profile, they do not cause significant temperature increases. We also point out that the radial structures in the potential vorticity closely resemble those in the pressure gradient profile. Actually, the minima in the normalised vorticity profile coincide fairly well with the maxima in the dust's surface density profile, a result that is further illustrated in the top-right panel of Figure~\ref{fig:fig_extra} at 220 $T_0$. We further observe a trend whereby the minima and maxima of the normalised vorticity profile are distributed at roughly equal intervals in log R. For instance, at 220 $T_0$, the radial minima in the normalised vorticity profile follow, approximately, a geometric sequence with a ratio of about 1.2 below 100 au, and about 1.4 above 100 au. Curiously, similar behaviour was reported in \citet{WFB20}, where several radial minima in the potential vorticity formed as a result of the intermittent migration of a massive planet in its disc. 

Finally, another reason for examining the normalised vorticity is that it is directly involved in the sufficient criterion for the onset of the Rossby-Wave Instability (RWI), which \citet{ChangYoudin24} formulated as $\kappa^2/\Omega^2 \lesssim 1/2$ for discs with negligible self-gravity. Although self-gravity could modify this criterion, since self-gravity impacts which modes of the RWI grow fastest \citep{lovelace2013}, it turns out that $\kappa^2/\Omega^2$ does not go well below 1/2, and no sign of the RWI is observed in our simulations (we come back to this point in the next section).

% -------------------------------------------
\subsection{From dust spirals to dust rings}
\label{subsec:change}
\begin{figure}
    \centering
       \includegraphics[width=0.93\hsize]{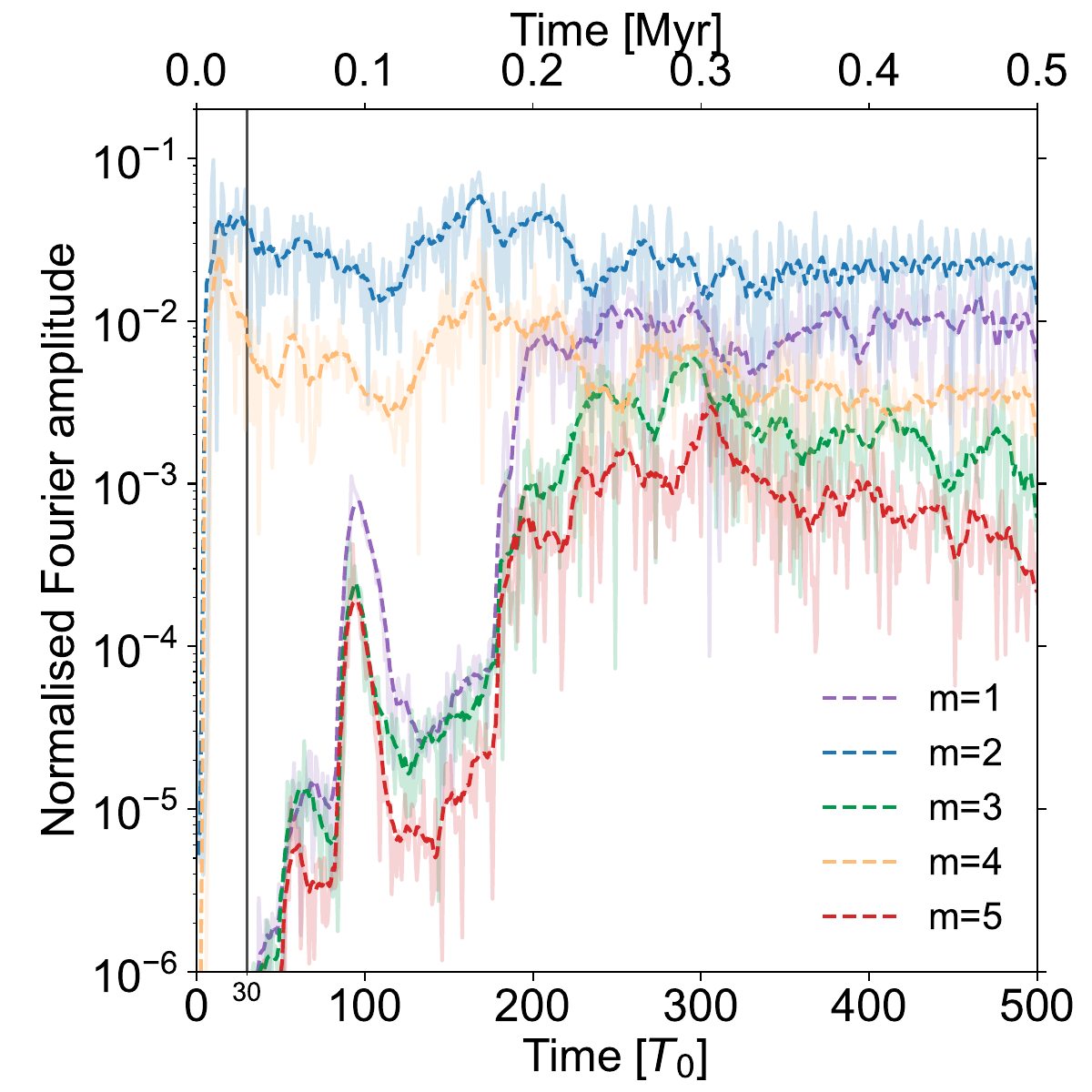}
    \caption{Temporal evolution of the modes $m=1$ to $m=5$ in the power spectrum of the gas surface density. The transparent solid curves show the instantaneous amplitudes, whereas the dashed curves show the same amplitudes, but smoothed over 10 $T_0$.}
    \label{fig:fig_mode}
\end{figure}

% -------------------------------------------
\noindent We wrap up this section with Figure~\ref{fig:fig_2Ddens}, which displays the gas and dust surface densities in Cartesian coordinates centred on the star. Each column of panels corresponds to a different time in the simulation. All the gas panels show spirals, but their contrast and amplitude decrease in time, as do the mass of the disc gas and its level of gravitoturbulence. A Fourier analysis of the gas surface density can help quantify this. The normalised amplitude of the modes (relative to the azimuthally-averaged density) is displayed as function of time in Figure~\ref{fig:fig_mode}. It shows that, quite early in the simulation, the mode with azimuthal wavenumber $m=2$ has the largest amplitude. Its normalised amplitude varies over time though it decreases overall. Interestingly, \citet{Huhn2026} also highlighted the prevalence of the $m=2$ mode in their simulations of non self-gravitating discs undergoing late infall, indicating that $m=2$ spirals may not necessarily be caused by GI. We note that the amplitude of the $m=1$ mode grows with time in our simulation but reaches a plateau from about 300 $T_0$. As already mentioned in \S\ref{subsec:numericsandbc}, our simulation does not show the reflex instability \citep{Crida2025b}, but other preliminary simulations did. 

Dust-wise, the lower panels in Figure~\ref{fig:fig_2Ddens} clearly show a morphological change in the dust surface density. At 77 $T_0$ , the dust surface density features large-scale spirals that coincide well with those in the gas surface density. At 190 $T_0$, while the dust density still closely resembles the gas density in the disc's inner parts, it is no longer the case in the disc's outer parts, where the Stokes number is largest. The dust gradually decouples from the gas and progressively forms rings. This is particularly clear for the two outermost dust rings located at approximately 200 au and 280 au, where prominent spirals are seen in the gas density. At 340 $T_0$, the dust density in the ring near 280 au has much decreased (see also the top-right panel in Figure~\ref{fig:fig_spacetime2}), but we still get multiple, well-separated and well-contrasted dust rings. The three most prominent rings are located near 52,110 and 220 au. There are two other rings that form below 30 au, but given their proximity to the inner edge of the computational domain, it is not clear that these rings are not artefacts of the boundary conditions and/or of wave reflections happening there. The ring near 110 au exhibits interesting substructures: on closer inspection, it can be seen as an asymmetrical ring with an undulating morphology shaped by a local two-armed spiral in the gas, and with an azimuthal contrast ratio in the dust density by a factor $\approx$5. Just inward of the ring, several blobs are seen, like tiny detached vortices. In the space-time diagram of the dust surface density (top-right panel in Figure~\ref{fig:fig_spacetime2}), these blobs correspond to the vertical zigzag shape located near 90 au, between about 300 and 400 $T_0$. The two other rings near 52 and 220 au also show azimuthal modulation in the dust density, with contrast ratios by approximately 25\% and a factor 2, respectively. Finally, at 460 $T_0$, we observe the same three distinct dust rings as before, with a few changes: the ring located near 110 au now exhibits strong azimuthal modulation (the azimuthal contrast ratio in dust density is now approximately a factor 100), and the ring located near 220 au has a wider radial extent (its FWHM, about 24 au, is very close to the local pressure scale-height). 

%FFFFFFFFFFFF
%FFFFFFFFFFFF

% =========================
\section{Radiative transfer calculations}
\label{sec:RT}
% =========================
\noindent Having found that millimetre-sized dust in late-infall-driven gravitoturbulent discs gradually changes from spirals to rings, we explore in this section how the gas and dust emission would look like if our model disc was observed at different epochs during its evolution. For this, we have post-processed the results of our hydrodynamical simulation to perform radiative transfer calculations with the public code \href{https://github.com/dullemond/radmc3d-2.0}{RADMC-3D} \citep{Dullemond2012}. The numerical setup of our radiative transfer calculations is described in \S\ref{subsec:RTsetup}, and their results are then presented in \S\ref{subsec:RTresults}.

%FFFFFFFFFFFF
\begin{figure*}
    \centering
       \includegraphics[width=\hsize]{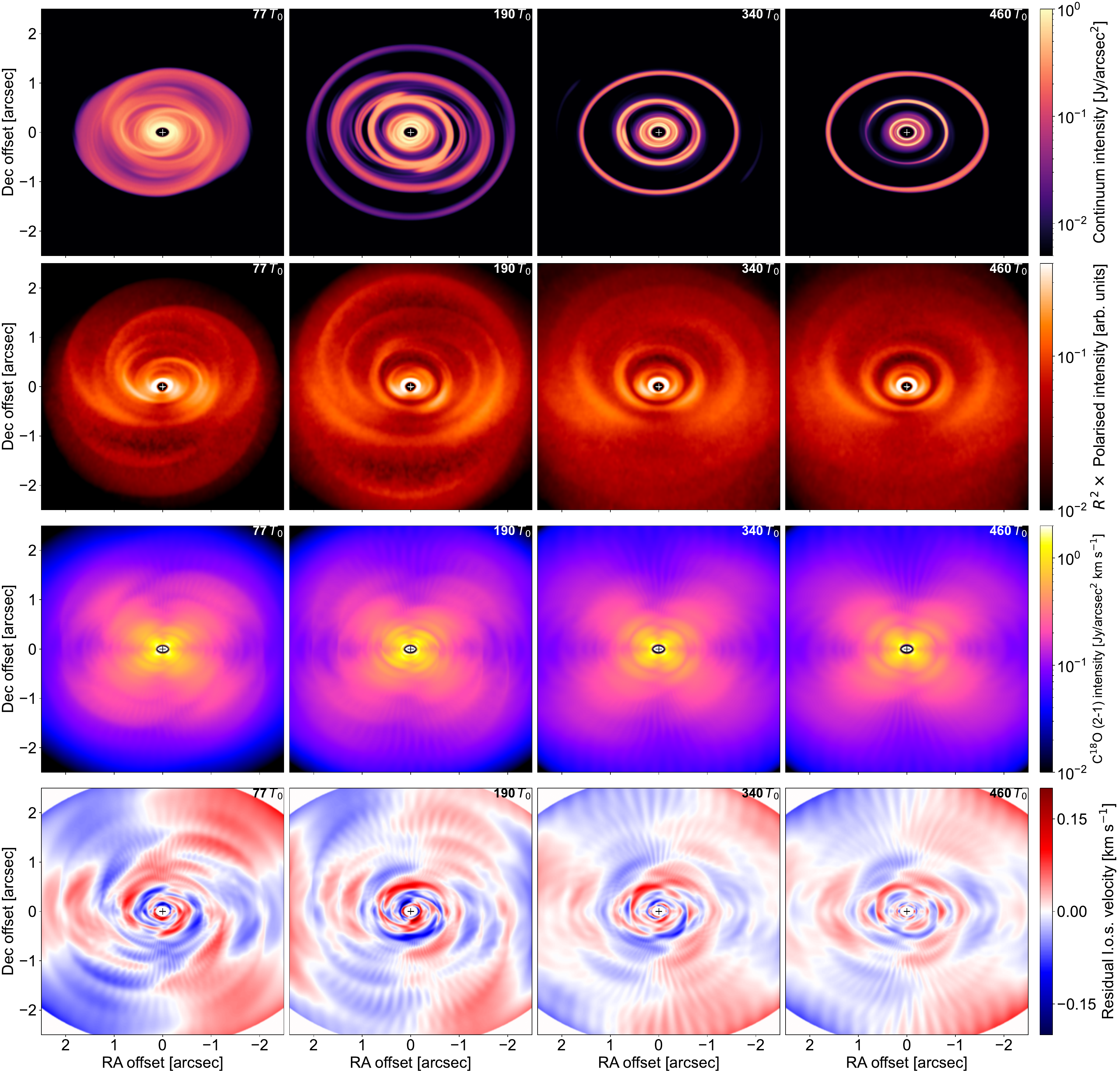}
    \caption{Synthetic observables obtained by post-processing our results of hydrodynamical simulation with RADMC-3D radiative transfer calculations. From top to bottom are displayed the dust's continuum intensity at $\lambda = 1.3$ mm, the dust's polarised intensity at $\lambda = 1.65$ $\mu$m, the C$^{18}$O (2-1) line intensity around $\lambda = 1.35$ mm, and the residual line of sight velocity. Polarised intensity images are scaled by the squared de-projected radial distance from the star to better visualise the contribution from the outer disc. Residual velocity images are derived from the moment 1 maps of the C$^{18}$O (2-1) line emission, from which a best-fit background rotation profile obtained using \texttt{eddy} was subtracted (see text). In each panel, time is shown in the upper-right corner and the star position is marked by a cross in the centre.}
    \label{fig:fig_RTmedley}
\end{figure*}
%FFFFFFFFFFFF

% -------------------------------------------
\subsection{Numerical setup}
\label{subsec:RTsetup}
% -------------------------------------------

%FFFFFFFFFFFF
\begin{figure*}
    \centering
        \includegraphics[width=\hsize]{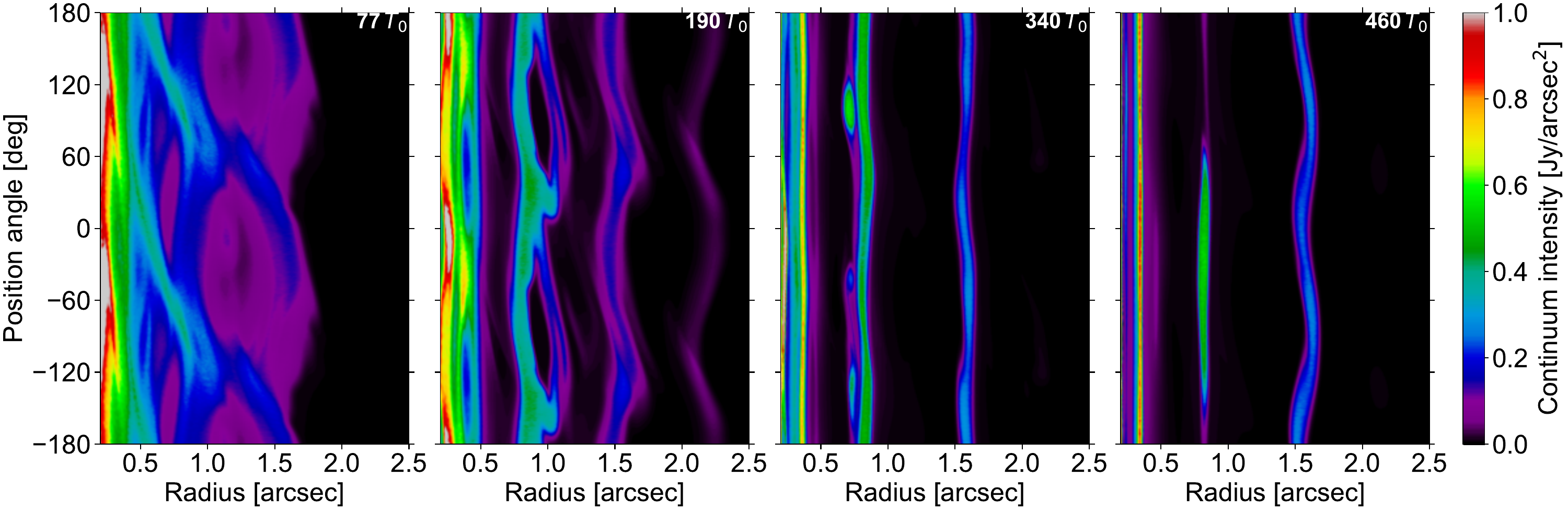}
    \caption{Same as the upper panels in Figure~\ref{fig:fig_RTmedley}, but now continuum intensity maps are displayed in the de-projected disc plane, in polar coordinates. The x-axis shows the de-projected orbital radius in arcseconds, and the y-axis shows the position angle (East of North) in degrees.
    }
    \label{fig:fig_extra_RT}
\end{figure*}
%FFFFFFFFFFFF

\noindent The public python program \href{https://github.com/charango/fargo2radmc3d}{\texttt{fargo2radmc3d}} was used to (i) post-process our results of hydrodynamical simulation so as to provide RADMC-3D with all necessary inputs for the radiative transfer calculations, (ii) call RADMC-3D, and (iii) produce synthetic images of the disc from the results of the radiative transfer calculations. Our disc is assumed to be located at 140 pc, to be inclined by 40 degrees relative to the sky plane, and its position angle is set to zero to facilitate the direct comparison between the synthetic images and the results of the hydrodynamical simulation. The details of above step (i) can be found in \citet{Baruteau2019} for dust radiative transfer calculations, and in \citet{Baruteau2021} for gas line radiative transfer calculations. Below we briefly summarise the main assumptions and parameters.
\\
\par\noindent{\it Dust continuum at 1.3 mm -- } To produce synthetic images of the dust continuum intensity at 1.3 mm wavelength, we have converted the surface density of the mm-sized dust fluid obtained in our hydrodynamical simulation into a volume density. This is done by assuming a Gaussian distribution for the dust volume density along the disc's vertical direction, with scale height $H_{\rm d} = H_{\rm g} (1 + {\rm St}/\alpha_{\rm z})^{-1/2}$ (\citealp{Dubrulle1995}, see also section 5.1 in \citealp{RendonRestrepo2025}). In the previous expression, $H_{\rm g}$ is calculated from the gas temperature assuming vertical hydrostatic equilibrium, St denotes the Stokes number, and $\alpha_{\rm z}$ is an alpha-viscosity supposed to mimic the effect of dust turbulent diffusion in the vertical direction. Using 3D shearing-box simulations, \citet{Riols2020} and \citet{Baehr2021} have shown that in gravitoturbulent discs, dust turbulent diffusion is much less efficient in the vertical than in the radial direction. Said differently, $\alpha_{\rm z}$ is lower than $\alpha_{\rm R} + \alpha_{\rm G}$, the sum of the disc-averaged alpha parameters associated with the Reynolds and gravitational stresses, by a factor ranging from a few to a few tens. To reflect this, we take $\alpha_{\rm z} \approx (\alpha_{\rm R} + \alpha_{\rm G})/10$. In practice, we set $\alpha_{\rm z}$ to $10^{-3}$ at 77 and 190 $T_0$, and to $2\times 10^{-4}$ at 340 and 460 $T_0$ (see Figure~\ref{fig:fig_discmass}). The 3D grid used by RADMC-3D spans $\pm3 H_{\rm g}$ about the disc midplane with 50 cells logarithmically spaced in colatitude. To account for the contribution of smaller dust to the continuum intensity, and to make up for the lack of multiple dust fluids in Dusty FARGO-ADSG, we have used the gas surface density to construct the mass volume density of another dust fluid modelling 10 $\mu$m-sized dust, assuming this time $H_{\rm d} = H_{\rm g}$. Dust particles of 10 $\mu$m size should have maximum Stokes numbers in the range $[10^{-5}-10^{-3}]$ between 50 and 300 au in our disc model, so it is reasonable to expect them to be well coupled to the gas. For this mock 10 $\mu$m dust fluid, the gas surface density was rescaled such that the total mass of the 10 $\mu$m fluid be approximately one third of that of the 1 mm dust fluid. Furthermore, the dust temperature is simply assumed to be that in the hydrodynamical simulation and to be independent of the disc's vertical coordinate $z$ (azimuthal variations in the dust temperature thus reflect heating at shocks due to spirals in the gas). Dust opacities are computed with Mie theory, assuming a mixed composition with 70\% water ices and 30 \% silicates \citep{Baruteau2019}. This yields an internal mass volume density for the dust of about 1.3 g cm$^{-3}$, which is the same value as in the hydrodynamical simulation. Our RADMC-3D calculations include anisotropic scattering, with a Henyey-Greenstein scattering phase function.
\\
\par\noindent{\it Dust polarised emission at 1.65 $\mu$m -- } Synthetic images of the dust polarised intensity are computed at 1.65 $\mu$m wavelength, which is assumed to arise from dust grains between 0.01 and 1 $\mu$m that are perfectly coupled to the gas. This small dust population is discretised into 6 size bins. For each bin, the mass volume density is calculated from the gas surface density assuming a size distribution in $s^{-3.5}$ and a total dust-to-gas mass ratio of $10^{-4}$ for this small dust population (we take $H_{\rm d} = H_{\rm g}$). The 3D grid used by RADMC-3D now spans $\pm5 H_{\rm g}$ about the disc midplane with 50 cells evenly spaced in colatitude. The dust temperature is again that in the hydrodynamical simulation, and we assume the same mixed composition as above (70\% water ices and 30 \% silicates).
\\
\par\noindent{\it C$^{18}$O (2-1) line emission -- }  We have also computed radiative transfer calculations of the C$^{18}$O J=2$\rightarrow$1 rotational line emission near 1.3 mm wavelength. The C$^{18}$O number density is obtained from the gas surface density rescaled by $3\times 10^{-8}$ to reflect a possible value for the C$^{18}$O-to-H$_2$ abundance ratio \citep[e.g.,][]{Zhang2021MAPS}, and a Gaussian vertical distribution is again assumed in the disc's vertical direction. For simplicity, effects of photodissociation and CO freeze-out onto dust grains are discarded. The 3D grid used by RADMC-3D spans $\pm5 H_{\rm g}$ about the disc midplane with 50 cells evenly spaced in colatitude. 3D cubes of the gas temperature and velocity are constructed assuming both fields are independent of $z$. The specific intensity of the line emission is computed in 101 channel maps covering $\pm$5 km s$^{-1}$ about the systemic velocity, which yields a spectral resolution of 0.1 km s$^{-1}$. Our RADMC-3D calculations use local thermodynamic equilibrium and the Leiden LAMBDA data base to compute the partition function for the level populations. Turbulent broadening arising from the gravitoturbulence is discarded.

% -------------------------------------------
\subsection{Results of calculations}
\label{subsec:RTresults}
% -------------------------------------------
\noindent Our results of radiative transfer calculations are displayed in Figure~\ref{fig:fig_RTmedley}. None of the synthetic images include noise or beam convolution: they should be regarded as {\it ideal} images in terms of angular resolution, sensitivity and spectral resolution. Gas synthetic images are also ideal in that they are derived from radiative transfer calculations in the gas only, thereby avoiding the need to subtract the dust continuum. The right ascension and declination offsets on the x- and y-axes span $\pm 2\farcs5$, which corresponds to a physical scale of 350 au: the size of the synthetic images is the same as that of the images showing the results of the hydrodynamical simulation in Figure~\ref{fig:fig_2Ddens}.

%FFFFFFFFFFFF
\begin{figure*}
    \centering
        \includegraphics[width=\hsize]{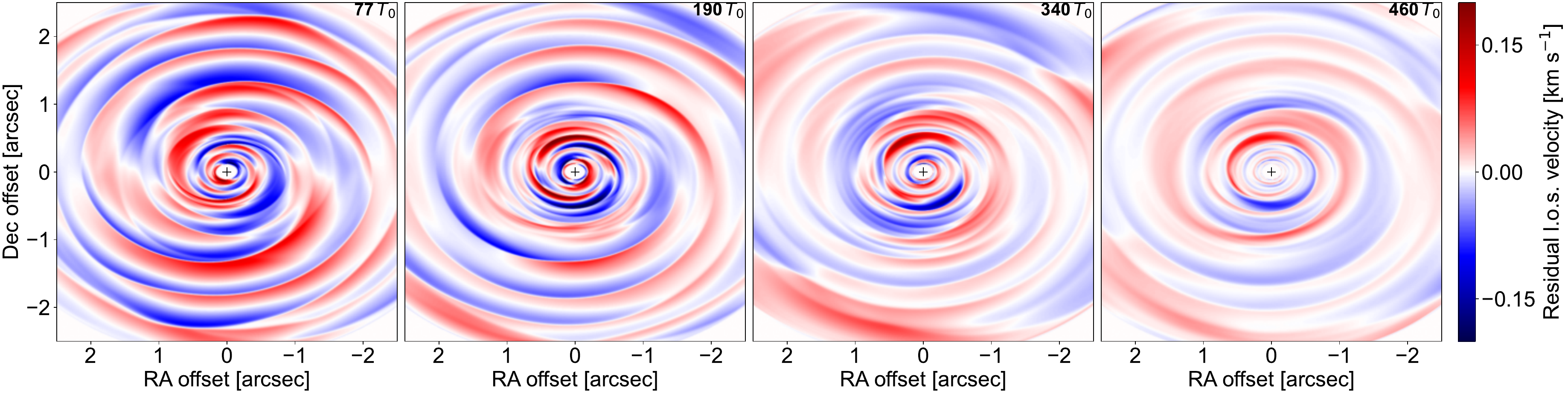}
    \caption{Residual line of sight velocity of the gas in the disc midplane, directly obtained from the velocity field in the hydrodynamical simulation.
    }
    \label{fig:fig_los_hydro}
\end{figure*}
%FFFFFFFFFFFF

% - - - - - - - - - - - - - - - - - -
\subsubsection{Dust continuum at 1.3 mm}
\label{subsubsec:continuum}
% - - - - - - - - - - - - - - - - - -
\noindent Images of the dust continuum intensity at wavelength $\lambda = 1.3$ mm are shown in the first row of panels in Figure~\ref{fig:fig_RTmedley}. De-projected images using polar coordinates and a linear colour scale can be found in Figure~\ref{fig:fig_extra_RT}. Notwithstanding the disc inclination, one easily recognises the structures of the dust surface density in the hydrodynamical simulation (compare with the lower panels in Figure~\ref{fig:fig_2Ddens}). This tells us that the continuum emission cannot be overly optically thick throughout the disc, which we have checked by computing maps of the optical depth via RADMC-3D (for instance, at 77 $T_0$, the optical depth exceeds about 3 only in the central $0\farcs5$). We will come back to this point in the penultimate paragraph of this section, when estimating dust masses from the integrated intensity. But most importantly, we see that the shape of the continuum intensity gradually changes from multiple spirals to multiple rings. 

The first image from the left at 77 $T_0$ is quite reminiscent of the 1.3 mm ALMA image of the Elias 2-27 disc \citep{LPerez2016, PanequeCarreno2021}, although the latter has a different inclination and position angle compared to our simulated disc. A similar morphology in the continuum images, characterised by multiple spirals, is obtained from the end of the infall episode at 30 $T_0$ up to approximately 100 $T_0$. After this time, spirals gradually give way to rings.

The second image at 190 $T_0$ corresponds to a transition time when the bright rings still show significant modulation in azimuth due to gas spirals crossing the dust rings. With a little imagination, the more spiral-like appearance of the disc's inner regions and the more ring-like appearance of the outer regions are vaguely reminiscent of the 1.3 mm ALMA image of the IM Lup disc \citep{Huang2018}.

The images at 340 $T_0$ and 460 $T_0$ show a disc with multiple concentric dark and bright rings in the continuum emission, which is broadly reminiscent of many observed discs with annular substructures \citep{Andrews2018,Cieza2021}. At 340 $T_0$, the asymmetrical dust ring near 110 au takes the form of a lopsided bright ring near $0\farcs8$ with an azimuthal contrast ratio in the intensity by about 20\%, and the blobs just inward of the dust ring show up as several crescent-shaped structures. The projected distance between the blobs and the ring is approximately $0\farcs1$ (see the third panel of Figure~\ref{fig:fig_extra_RT}), which means that the blobs and the ring would remain distinguishable if our synthetic image were convolved with a realistic ALMA beam at this wavelength. The crescent-shaped structure located near 9 o'clock bears strong resemblance to that seen in the 1.3 mm ALMA image of the HD 163296 disc \citep{Isella2018}. Recall from \S\ref{subsec:change} that the aforementioned blobs appear as a vertical zigzag pattern in the space-time diagram of the dust's surface density in the top-right panel of Figure~\ref{fig:fig_spacetime2}. From this we can deduce that the crescent-shaped structures live for about 100 $T_0$ in our model (i.e., approximately 0.1 Myr). At 460 $T_0$, the dust ring near 110 au now appears as a lopsided bright ring with an azimuthal contrast ratio in the intensity between approximately 5 and 10 (see the fourth panel in Figure~\ref{fig:fig_extra_RT}). It is still present at the end of the simulation, meaning that it has lived for at least 100 $T_0$. The large azimuthal extent of this ring makes it reminiscent of lopsided rings observed at mm wavelengths in the AB Aur disc \citep{Tang2012, Fuente2017, Riviere2024} or in the MWC 758 disc \citep{Dong2018}.

Note that dark rings do not get too dark in our synthetic images due to the inclusion of the mock 10 $\mu$m dust fluid coupled to the gas (see \S\ref{subsec:RTsetup}), which contributes to a faint background emission in the images. This and optical depth effects, due in part to the disc inclination, imply that the radial and the azimuthal contrasts of the rings are smaller in the synthetic continuum images than in the dust's surface density images from the hydrodynamical simulation. 

We further point out that the total integrated flux at 1.3 mm is respectively about 1.18, 0.92, 0.58 and 0.41 Jy from left to right in the top panels of Figure~\ref{fig:fig_RTmedley}. Assuming optically thin emission, a dust temperature of 20 K and an absorption opacity of 2.3 cm$^2$/g (as typically assumed in disc demographics studies; see for instance \citealp{Bhowmik2026}), we would estimate total dust masses from our synthetic images of, respectively, 653, 509, 321 and 227 Earth masses. These values should be compared with the actual values in our simulation, which are respectively 522, 457, 439 and 436 Earth masses. The fact that the estimated masses are greater than their actual values at 77 $T_0$ and 190 $T_0$ is a combination of the absorption opacity being higher in our calculation ($\approx$ 4.0 cm$^2$/g) and of the flux primarily originating from the inner regions of the disc, where the temperature is higher than 20 K. The fact that the estimated masses are indeed lower than their actual values at 340 $T_0$ and 460 $T_0$, which remains true even for opacities and temperatures closer to those in our calculations, can be explained by optical depth effects. Still, it is interesting to note that the estimated and actual masses do not differ significantly in our calculations, despite the numerous uncertainties involved in determining a dust mass from the total flux of continuum emission.

Finally, we note that, in our model, late infall increases the mass of the disc gas quite substantially (by a factor 3, see Figure~\ref{fig:fig_discmass}) without any dust being added during the process. This is why the mass ratio between the mm-sized dust and gas, initially set at 1\%, remains between approximately 0.3\% and 0.37\% from the moment when infall ceases until the end of the simulation (see \S\ref{subsec:spacetime}). However, infall is also likely to add copious amounts of {\it a priori} small dust to the disc. This small dust will gradually grow, thereby increasing the mass in mm-sized solids and influencing the resulting continuum images. How efficient dust growth is in a gravitoturbulent disc like ours is a highly relevant question that deserves a specific study \citep[e.g.,][]{Booth2016}. As long as dust drag on the gas can be neglected, the mass of mm-sized solids can be treated as a free parameter in dust radiative transfer calculations. In particular, we have verified that scaling up the density of the mm-sized fluid  by a factor of three has a minor impact on the continuum images shown in Figure~\ref{fig:fig_RTmedley}.

% - - - - - - - - - - - - - - - - - -
\subsubsection{Dust polarised emission at 1.65 $\mu$m}
% - - - - - - - - - - - - - - - - - -
\noindent The second row of panels in Figure~\ref{fig:fig_RTmedley} displays synthetic images of the polarised intensity at $\lambda=1.65$ $\mu$m. At 77 and 190 $T_0$, the polarised intensity shows distinct spirals wrapping up on the upper surface of the disc, and at least one spiral is visible in the lower surface. As time increases, it is clear that the spirals in the disc surfaces gradually fade away, which is consistent with the gas surface density panels in Figure~\ref{fig:fig_2Ddens}. The polarised intensity images at 340 $T_0$ and 460 $T_0$ are very similar and allow to see the gaps in the gas density that are sandwiched between the three outermost bright rings in the continuum intensity, that is near $0\farcs6$ (80 au) and $1\farcs2$ (160 au). We point out that the last two images show similarities to the polarised intensity image of the HD 34282 disc obtained with SPHERE \citep{deboer2021}.

We note the presence of a rather extended darker area located to the south in the images, and which fades over time. We attribute this feature to optically thin dust emission. As we have seen in \S\ref{subsec:spacetime}, the outer disc gas gradually expands outwards, and the local gas density increases with time (see top-left panel in Figure~\ref{fig:fig_spacetime2}). Given our total opacity at 1.65 $\mu$m, the small (sub)-$\mu$m dust emission in the disc's outermost regions is optically thin. The dark band that is visible at declination offsets around -1$\farcs$2 at 77 $T_0$ thus marks an area where the dust density is smaller, and the line of sight emission thus dimmer. This band appears at larger (negative) declination offsets at 190 $T_0$ due to the disc's outward expansion, and it is no longer visible at 340 and 460 $T_0$ owing to the local increase in gas density. We also note the presence of a dark cone centred around the 6 o'clock position in all four images, which we attribute to the dust's scattering phase function.

In a large census of near-IR images of protoplanetary discs, \citet{Garufi2026} have shown that discs with detected ambient material have no rings in scattered light, but half of them exhibit spirals. The evolutionary sequence shown in this section is consistent with this result: our synthetic near-IR images reveal spiral structures during the final episode of late infall and shortly after it ceases, whereas they ultimately show rings long after infall has stopped.

% - - - - - - - - - - - - - - - - - -
\subsubsection{C$^{18}$O (2-1) line emission}
\label{subsubsec:c18o}
% - - - - - - - - - - - - - - - - - -
%FFFFFFFFFFFF
\begin{figure*}
    \centering
        \includegraphics[width=\hsize]{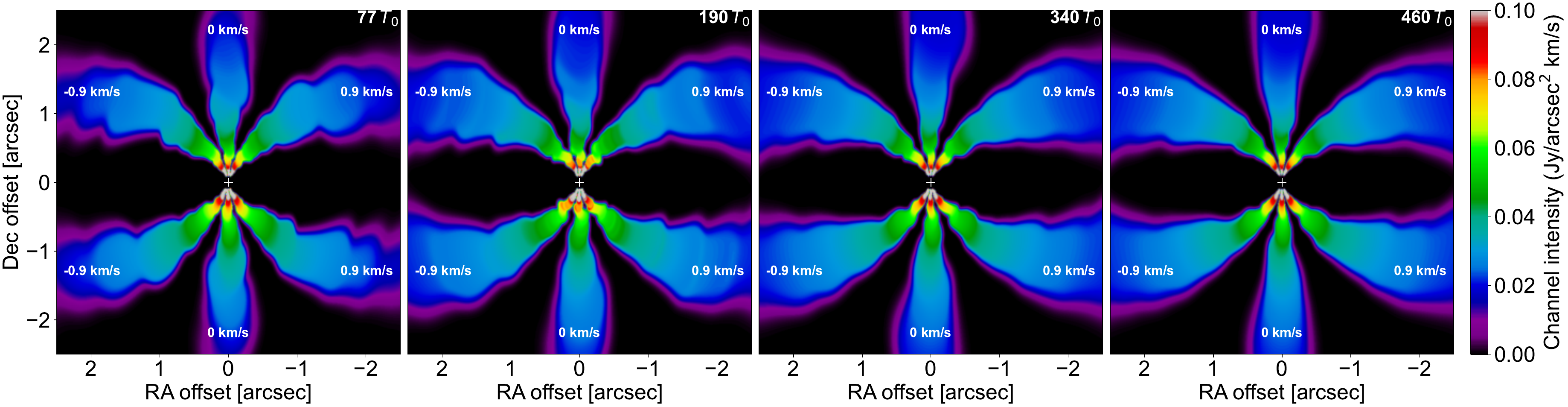}
    \caption{Selected channel maps of the C$^{18}$O (2-1) line emission, corresponding to velocity shifts relative to the systemic velocity of 0 and $\pm$0.9 km/s.
    }
    \label{fig:fig_extra_RT_wiggle}
\end{figure*}
%FFFFFFFFFFFF
\noindent Moment 0 maps of the C$^{18}$O J=2$\rightarrow$1 rotational line emission follow in the third row of panels in Figure~\ref{fig:fig_RTmedley}. At first sight, the four panels look very similar: the main structure that stands out of the maps is a four-lobed pattern that comes about because of optical depth effects, and which imply local minima in the integrated intensity along the disc's minor and major axes \citep{Rosenfeld2013,Baruteau2021}. But a closer comparison between the panels allows to distinguish additional spiral structures in the outer disc parts in the moment 0 map at 77 $T_0$, and to a lesser extent at 190 $T_0$ (spirals are better seen around the disc's major axis, located at 0 declination offset). We have checked that these spiral structures actually overlap well with the outermost spirals seen in the gas surface density panels in Figure~\ref{fig:fig_2Ddens}, modulo the disc inclination. At 340 and 460 $T_0$, the gap near $0\farcs6$ (80 au) is visible, that near $1\farcs2$ (160 au) is far less so.

To gain insight into the kinematic signatures of our disc model, we have calculated moment 1 maps of the line emission, which are displayed in the upper panels of Figure~\ref{fig:fig_mom1maps}, in \S\ref{sec:extrafigures} of the Appendix. To determine the residual velocity along the line of sight (hereafter, l.o.s.), a background velocity field must be subtracted from the moment 1 maps. To do so, we have used \href{https://github.com/PlanetFormationLab/eddy}{\texttt{eddy}} \citep{eddy} to compute a best-fit model for the disc's background velocity field, taking into account the radial mass distribution of the disc, denoted by $M_{\rm disc}(R)$. In \texttt{eddy}, this is done by assuming that the de-projected background velocity (i.e., the disc's unperturbed velocity in the disc plane) reads 
\begin{equation}
v_{\varphi}(R) = \sqrt{ \frac{G \{ M_{\star} + M_{\rm disc}(R) \} }{R} },
\end{equation}
with 
\begin{equation}
M_{\rm disc}(R) = M_{\rm disc} \times \frac{R^{2-\varpi} - R_{\rm min}^{2-\varpi}}{R_{\rm max}^{2-\varpi} - R_{\rm min}^{2-\varpi}},
\end{equation}
where $R_{\rm min}$ and $R_{\rm max}$ are the minimum and maximum radii of the disc. Just like the disc mass $M_{\rm disc}$, the exponent $\varpi$ or the star mass $M_{\star}$, these are parameters (among others) fitted by \texttt{eddy} using an MCMC calculation. The resulting residual velocity maps are displayed in the fourth row of panels in Figure~\ref{fig:fig_RTmedley}. Spirals are clearly visible in the images, and their contrast gradually decreases over time. For comparison purposes, we have also produced residual velocity maps by subtracting instead the (de-projected) azimuthal velocity field directly obtained from the hydrodynamical simulation. These maps are shown in the lower row of panels in Figure~\ref{fig:fig_mom1maps}, which are in very good agreement with those obtained by using \texttt{eddy}. Discarding the disc mass in the background velocity field would result in a large-scale background signal in the residual velocity maps, exhibiting an east/west asymmetry (analogous to an $m=1$ mode). 

Alternatively, an easier and faster approach is to compute the residual l.o.s. velocity directly from the velocity field of the hydrodynamical simulation, instead of using the velocity field derived from the moment 1 maps of the gas emission. This is what is shown in Figure~\ref{fig:fig_los_hydro}. Since our hydrodynamical simulation is 2D, the residual l.o.s. velocity would correspond to that arising from the disc midplane. It would therefore only be comparable to observed residual velocity maps if the observed gas emission originated solely from the disc midplane. We see that, in the central arcsecond, the panels in Figure~\ref{fig:fig_los_hydro} bear a strong resemblance to the synthetic residual velocity maps shown in the lower panels of Figure~\ref{fig:fig_RTmedley}. However, unlike the synthetic maps, the maps obtained directly from the hydrodynamical simulation show large-scale spirals in the entire disc. Their amplitude gradually decreases with time, in agreement with what precedes. At 77 and 190 $T_0$, the residual l.o.s. velocity reaches typically 0.2 km s$^{-1}$. These amplitudes are similar to those of the spirals detected in the centroid velocity residuals for several sources of the exoALMA Large Programme \citep[][see their figure 4]{Fukagawa2026}. Also, these amplitudes can be compared with the sound speed in our disc model, which is in the range $[0.25-0.4]$ km s$^{-1}$ between 50 and 350 au. The residual velocities that we obtain thus reach near sonic amplitudes, which is consistent with previous works \citep[e.g.,][]{BethuneGI2021}. 

Finally, the large-scale spirals imply that we get wiggle signatures in our channel maps of the C$^{18}$O J=2$\rightarrow$1 line emission that resemble those obtained by \citet{Hall2020} in their 3D simulations of cooling-driven GI discs. This is demonstrated in Figure~\ref{fig:fig_extra_RT_wiggle}. At each time, three channel maps are shown, corresponding to velocity shifts with respect to the systemic velocity of -0.9, 0 and 0.9 km/s. Since the amplitude of the spirals (in perturbed density, in perturbed velocity) decrease in time, so does the amplitude of the wiggles in channel maps. It would be interesting to investigate whether infall- and cooling-driven GI discs would have wiggle signatures that are observationally distinguishable, and compare them to infall-driven wiggle signatures obtained without GI \citep{Calcino2026}. In an infall-driven GI disc, it would also be interesting to quantify the wiggle amplitude and its correlation with the infall rate and/or the alpha viscosity associated to gravitoturbulent transport.

% =========================
\section{Concluding remarks}
\label{sec:conclusion}
% =========================
\noindent The original motivation of this work lies in the ALMA continuum observations of the disc around the young star Elias 2-27 \citep{LPerez2016, PanequeCarreno2021}, which is one of the few discs where the continuum emission in the (sub-)millimetre exhibits spiral structures \citep[see also][]{Huang2018, Bae2023}. Do we expect millimetre-sized dust to form spirals, just like the gas? And which physical mechanism is responsible for the spirals? Several works have addressed these questions, starting from \citet{Tomida2017} and \citet{Meru2017}. These two studies examined the possibility that spirals are induced by gravitational instability (GI). The present study aims at revisiting the long-term dynamics of millimetre dust in discs where GI drives the formation of large-scale spirals in the gas, without causing fragmentation. We have examined the case where GI sets in through a late episode of mass infall onto the disc. By episode, we mean that the infall is assumed to have a limited timespan (30 kyr in our model). We have carried out two-dimensional hydrodynamical simulations modelling gas and dust with a simplified treatment of infall. Although a number of simulations have been performed in the course of this work, we have chosen to present only one.

During infall, GI sets in and the disc develops a gravitoturbulent state characterised by the continuous formation of large-scale spirals in the gas. Once infall ceases, the disc mass, the amplitude of the gas spirals, and the level of gravitoturbulence gradually diminish, albeit non-monotonically. Shortly after the termination of infall, the millimetre dust is still well coupled to the gas, except in the disc's outermost parts where gas drag is most effective, and the dust surface density features large-scale spirals. Even long after infall has ceased, two-armed spirals remain persistently excited by GI in specific regions of the disc. It implies the gradual formation of radial structures in the disc gas that can be regarded as zonal flows, or as fossilised remnants of shocks due to the GI spirals. These zonal flows are long-lived maxima in the radial pressure gradient, and they progressively trap millimetre-sized dust. It is actually the main result of this work, which can be rephrased as follows: millimetre dust in infall-driven gravitoturbulent discs gradually evolves from multiple spirals to multiple rings. Said in another way: once late infall has ceased, a massive self-gravitating disc may gradually evolve from a disc with spirals in the continuum, like Elias 2-27, IM Lup or WaOph 6, to a disc with multiple dark and bright rings in the continuum, like HL Tau, HD 163296 or MWC 758.

Another important consequence is that, in near-IR scattered light, the disc initially exhibits multiple spirals during infall or shortly after it ceases, before ultimately displaying multiple rings. This result is consistent with the large sample of near-IR discs of \citet{Garufi2026}, where it is found that about half of the discs with ambient material (suggestive of possible infall) have spirals, but none of them have rings. As we have shown for the C$^{18}$O (2-1) emission, the persistent excitation of GI-driven spirals causes wiggles in channel maps, and implies that the residual line of sight velocity derived from line emission features large-scale spirals.

Previous works have highlighted the formation of dust rings in gravitationally unstable discs. This is the case, for example, with discs subject to the secular gravitational instability \citep[e.g,][]{Takahashi2016}, for which the 2D simulations of \citet{Pierens2021} showed that the rings are prone to secondary instabilities. This is also the case with the 3D shearing-box simulations of cooling-driven gravitoturbulent discs in \citet{Baehr2022}, where rings were observed for dust particles marginally coupled to the gas (see their figure 6), or with the 3D simulations of warped gravitoturbulent discs in \citet{Rowther2022}. Also, \citet{Vorobyov2024} carried out 2D gas+dust hydrodynamical simulations to model the formation of a protostar and its protoplanetary disc via the gravitational collapse of a pre-stellar core. They have highlighted the formation of a wide dust ring which coincides with the formation of a region in the disc gas inward of 1 au where the stress level is much reduced (a kind of gravitationally dead zone). This work was followed up by that of \citet{Matsukoba2025}, which showed the late formation of multiple dust rings via shocks driven by the inner wake of a distant clump formed by GI. This scenario actually bears analogies with the formation of multiple gaps in the gas by a planet in its disc via its inner wake \citep[e.g.,][]{Bae2017,Dong2018multiplegaps}.

Other mechanisms can form multiple dust rings in discs. A first class of mechanisms have in common the formation of spirals in the gas. These include: a low-mass planet with no or little migration \citep[e.g.,][]{Zhang2018,Perez2019}, a fast-migrating massive planet \citep{WFB20, Weiskopf2026}, several planets \citep[e.g.,][]{Dipierro2015}, the generation of a one-armed spiral by slowly growing eccentric modes and rapid cooling \citep{Li_ring_2021}, and the generation of multiple spirals by shadowing effects arising from a misaligned inner disc \citep{Ziampras2025, Zhang2025} or from external photo-evaporation \citep{Ziampras2026}. A second class of mechanisms operate in axisymmetric discs. These include zonal flows driven by magnetised winds \citep{Riols2020}, and self-induced dust traps driven by dust growth and drag onto the gas \citep{Gonzalez2017}. The interested reader will find more mechanisms in the review of \citet{Bae2023}. 

Our scenario thus adds to the many mechanisms capable of generating multiple dust rings in discs. In our opinion, the key questions here are: how can we distinguish between all these mechanisms? What are their individual signatures? With this in mind, we have produced synthetic observables from our results of hydrodynamical simulation for the main channels of disc observations: continuum emission in the millimetre, polarised intensity in the near-infrared, line emission in the millimetre and inference of the disc kinematics. Large-scale spirals across the disc in the residual line of sight velocity, together with indicators of a massive disc, are likely to be distinctive features of the formation of multiple dust rings via late infall-driven GI.

Our work leaves several open questions for future works. What is the long-term evolution of dust rings formed by infall-driven GI? How do the rings evolve in the presence of magnetised winds, of MHD turbulence, or if there were other subsequent episodes of mass infall? How do the number and position of the rings depend on the rate and the radial location of mass infall? What if the infall is not axisymmetric and/or torques the disc? Can infall-driven dust rings lead to planetesimal formation, or even to planet formation without need for fragmentation? How would our results compare to those of three-dimensional hydrodynamical simulations? Are there discs for which dust and gas observations can be simultaneously explained by a past episode of late infall-driven GI?

%=======================%
\section*{Acknowledgments}
%=======================%
\noindent CB and JFG acknowledge support from the "Programme National Astronomie-Astrophysique" of CNRS/INSU co-funded by CEA and CNES. SP acknowledges support from FONDECYT 1231663, ANID--Millennium Science Initiative Program NCN2024\_001 and ANID FIUF137139-USACH (CIRAS-AI). FM acknowledges support from The Royal Society Dorothy Hodgkin Fellowship. The simulations and analysis presented in this article were carried out on the CALMIP Supercomputing Centre of the University of Toulouse. We thank Josh Calcino, Antonio Garufi, and François Ménard for stimulating discussions. We thank the two reviewers for their careful reading of our manuscript, as well as for their very useful and constructive comments.
%=======================%

\begin{appendix}

% =========================
\section{A - A critical beta parameter for the onset of infall-driven fragmentation}
\label{sec:beta_infall}
% =========================
\noindent At the beginning of \S\ref{sec:resultshydro}, we have proposed that the onset of infall-driven fragmentation could be described by a critical value of the dimensionless timescale $\beta_{\rm in} = \Omega \times \Sigma / \dot{\Sigma}_{\rm in}$, which describes the rate at which the disc's surface density increases due to mass infall. By analogy with the well-known parameter $\beta_{\rm cool}$ for cooling-driven fragmentation, we have proposed that infall-driven fragmentation would be triggered when $\beta_{\rm in} \lesssim 3-4$. This appendix aims to provide more insight into this proposal.

Firstly, it can be compared with the criterion of \citet{Kratter2010}, who introduced two dimensionless parameters to describe infall: $\Gamma = \dot{M}_{\rm in} \,\Omega^{-1} (M_{\star}+M_{\rm disc})^{-1}$ and $\xi = G \dot{M}_{\rm in} \,c^{-3}_{\rm s}$, where all quantities have already been defined. Using 3D simulations of isothermal discs undergoing self-similar accretion from a pre-stellar core, \citet{Kratter2010} found infall-driven fragmentation when $\Gamma \lesssim \xi^{5/2}/850$. Applying this criterion to our disc model, and considering that $c_{\rm s}$ in the expression for $\xi$ is the adiabatic sound speed rather than the isothermal one, we would expect fragmentation for $\dot{M}_{\rm in} \gtrsim 10^{-5} M_{\odot}\,{\rm yr}^{-1}$ or, locally, for $\beta_{\rm in} \lesssim 10$ (with little overall dependence on $M_{\rm disc}$). This yields a critical value of $\beta_{\rm in}$ for infall-driven fragmentation that is consistent with our proposal, albeit slightly larger.

Another way of looking at the critical $\beta_{\rm in}$ for fragmentation is as follows: fragmentation will occur via infall if, locally, $\dot{\Sigma}_{\rm in} \gtrsim \dot{\Sigma}_{\rm max,\,GI}$, the maximum possible rate that gravitoturbulence can provide. We take the latter to be $\dot{\Sigma}_{\rm max,\,GI} \approx \dot{M}_{\rm max,\,GI} / \{\pi H^2\}$, that is the maximum mass accretion rate through gravitoturbulence over a surface set by the local pressure scale-height $H$. Assuming that mass transport induced by gravitoturbulence can be regarded as equivalent to a steady-state diffusion process (which is not necessarily the case, \citealp{BP1999}), $\dot{M}_{\rm max,\,GI} = 3\pi H^2 \Omega \Sigma \alpha_{\rm max,\,GI}$ with $\alpha_{\rm max,\,GI}$ the maximum alpha turbulent viscosity that gravitoturbulence can provide (not yet specified, see below). The inequality $\dot{\Sigma}_{\rm in} \gtrsim \dot{\Sigma}_{\rm max,\,GI}$ then implies $\beta_{\rm in} \lesssim 1 / \{3\alpha_{\rm max,\,GI}\}$ for infall-driven fragmentation to occur. Now, for $\alpha_{\rm max,\,GI}$, \citet{RendonRestrepo2026b} has shown that the maximum alpha turbulent viscosity provided by gravitoturbulence in non-fragmenting simulations of cooling-driven GI discs is $\alpha_{\rm max,\,GI} \approx 0.06$ (see the top panel of his figure 3; interestingly, this value is close to that found in \citealp{Rice2005}). Assuming the same value of $\alpha_{\rm max,\,GI}$ applies to infall-driven GI discs, the condition $\beta_{\rm in} \lesssim 1 / \{3\alpha_{\rm max,\,GI}\}$ becomes $\beta_{\rm in} \lesssim 5-6$, which is indeed close to our proposal. An $\alpha_{\rm max,\,GI}$ of $\approx$ 0.1 \citep[]{Longarini2026review} would yield $\beta_{\rm in} \lesssim 3-4$.

% =========================
\section{B - Additional figures}
\label{sec:extrafigures}
% =========================
\noindent In this section, we briefly introduce figures that complement those presented in \S\ref{sec:resultshydro} and \S\ref{sec:RT}.
\medskip
\par
\begin{figure*}[!h]
    \centering
        \includegraphics[width=0.32\hsize]{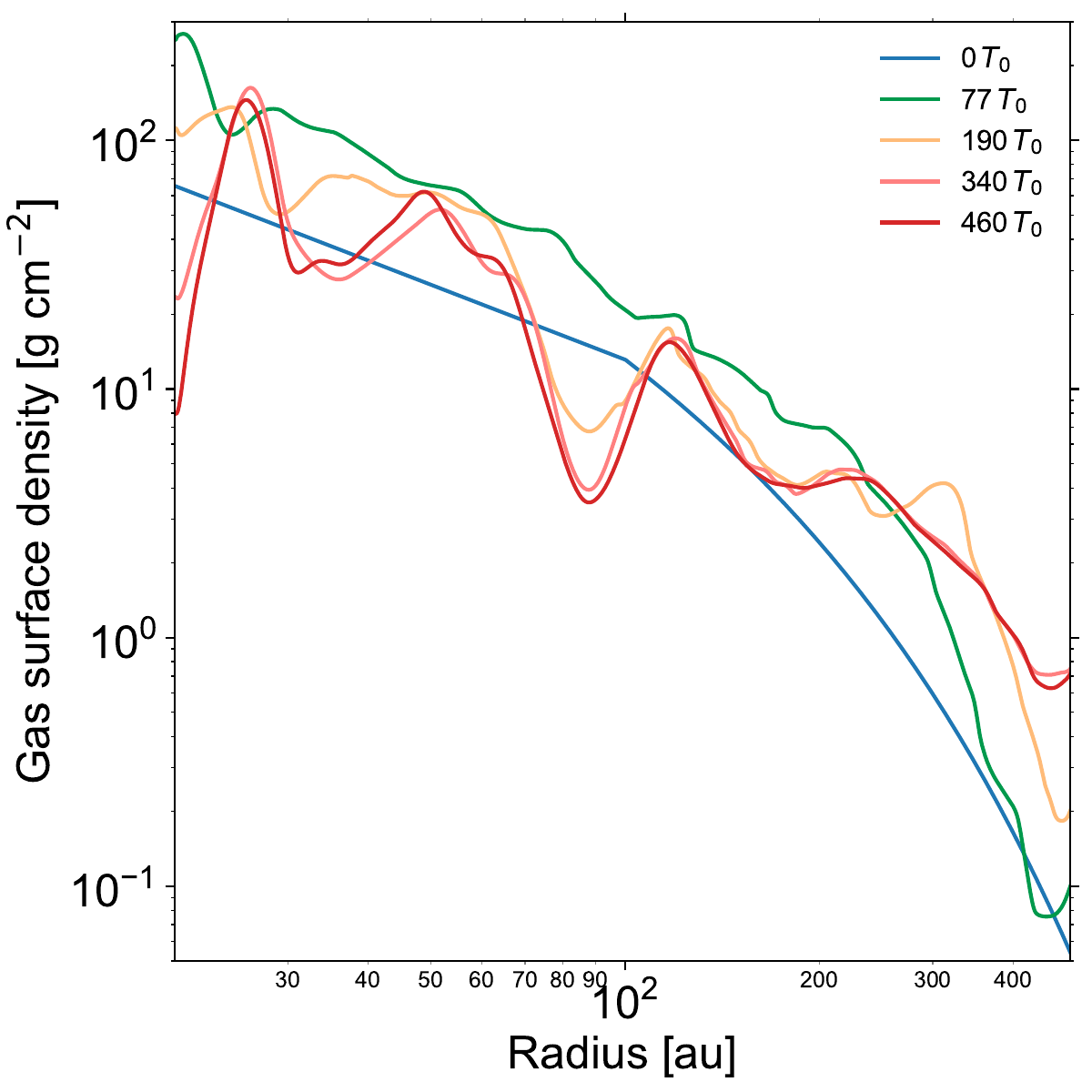}
         \includegraphics[width=0.32\hsize]{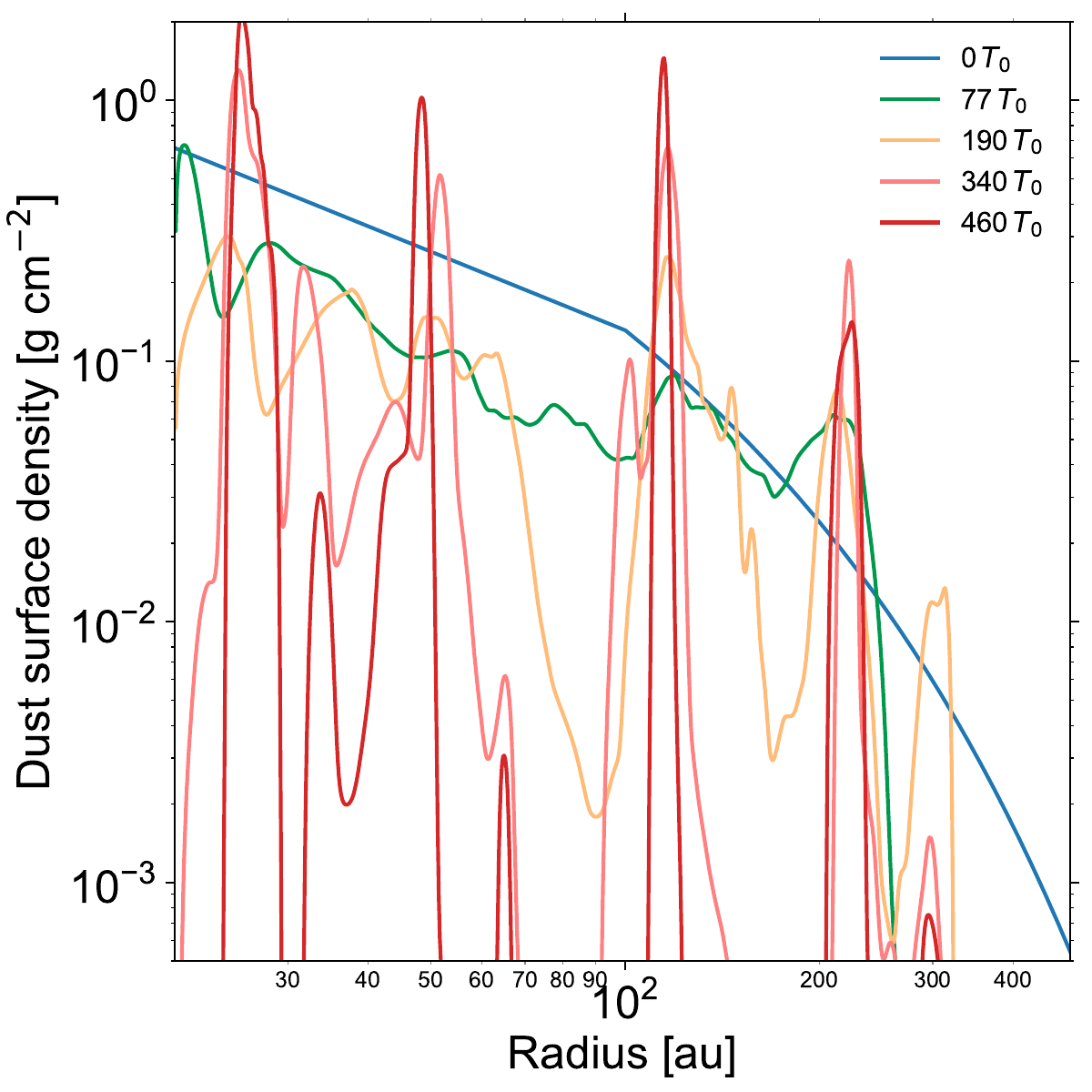}
	\includegraphics[width=0.32\hsize]{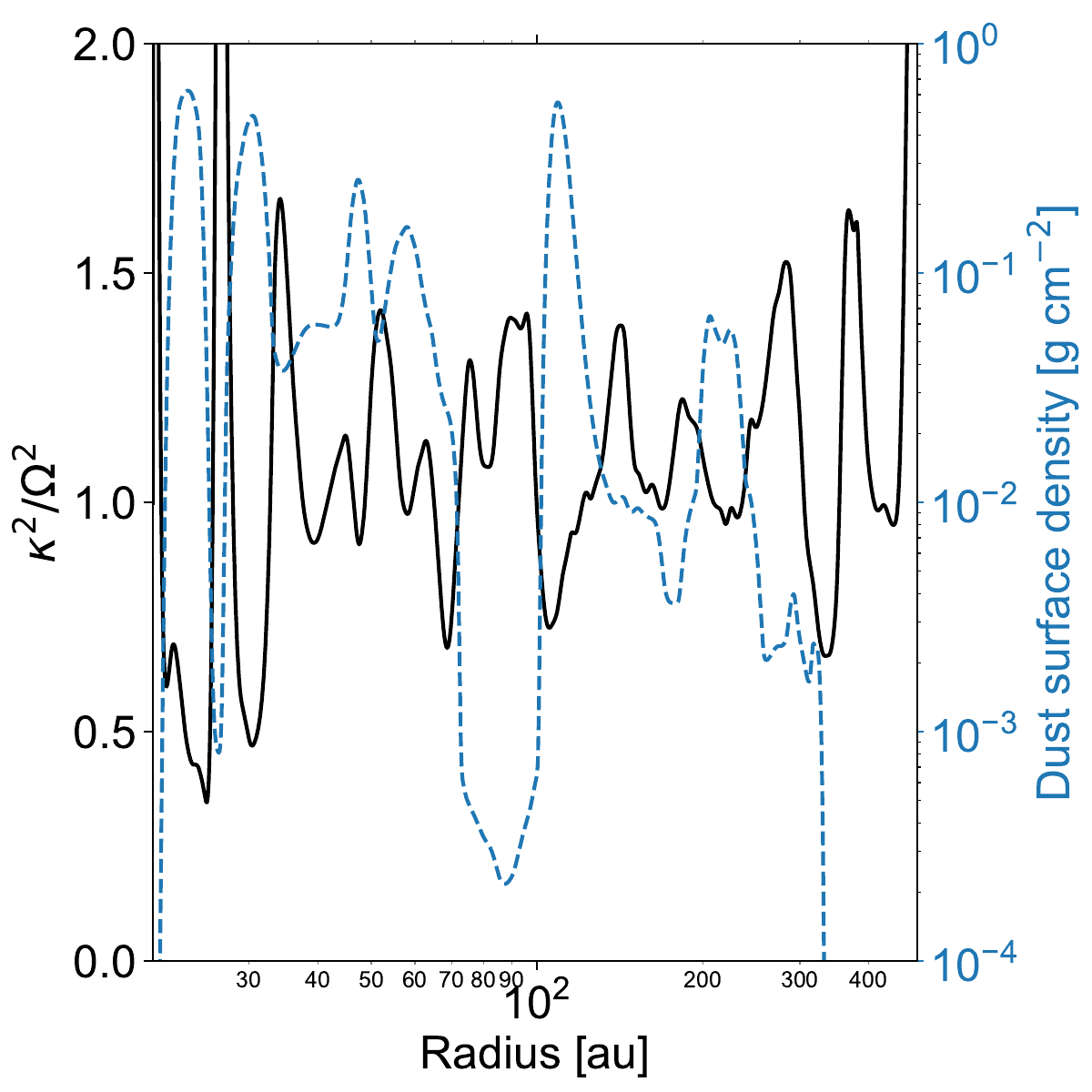}
        \includegraphics[width=0.32\hsize]{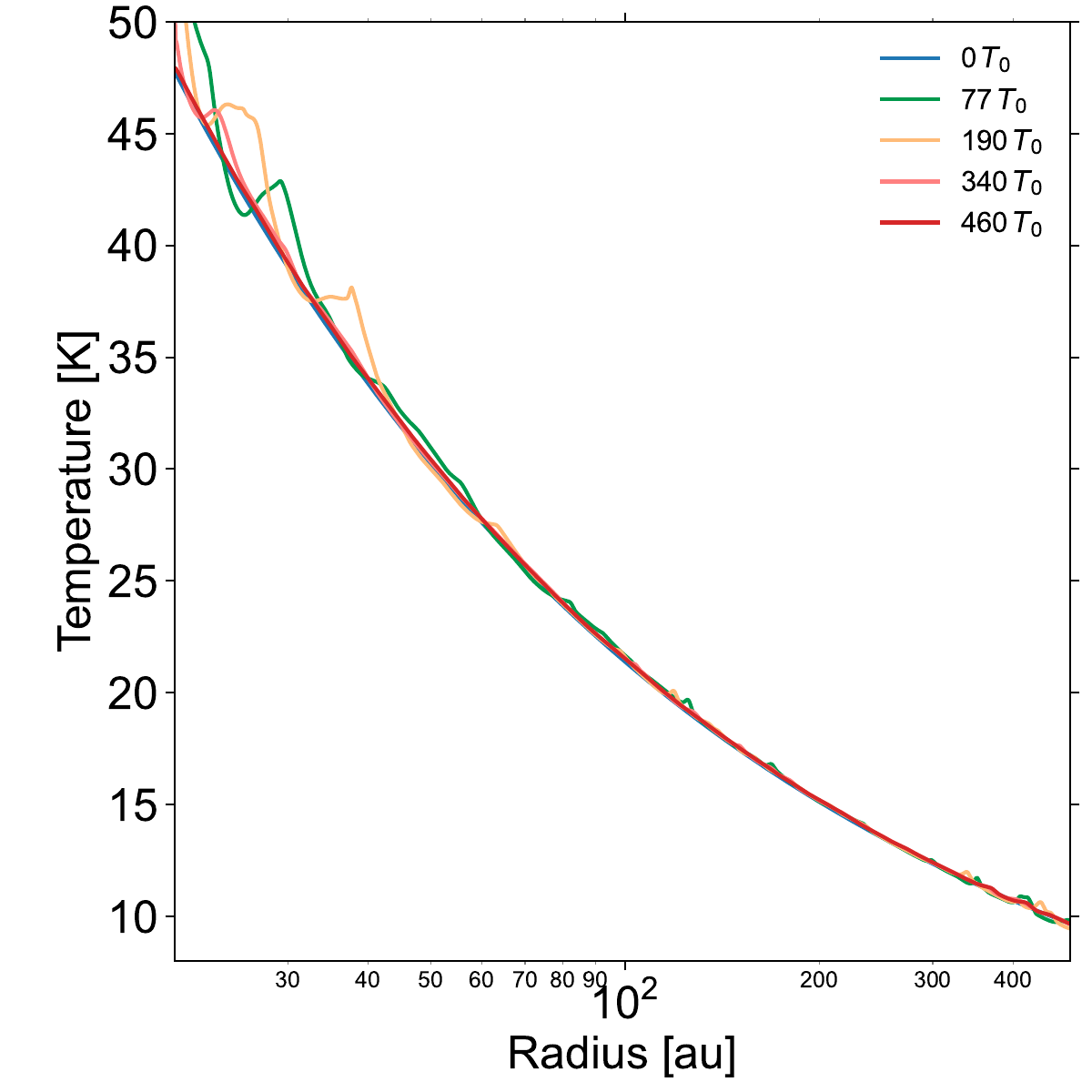}
        \includegraphics[width=0.32\hsize]{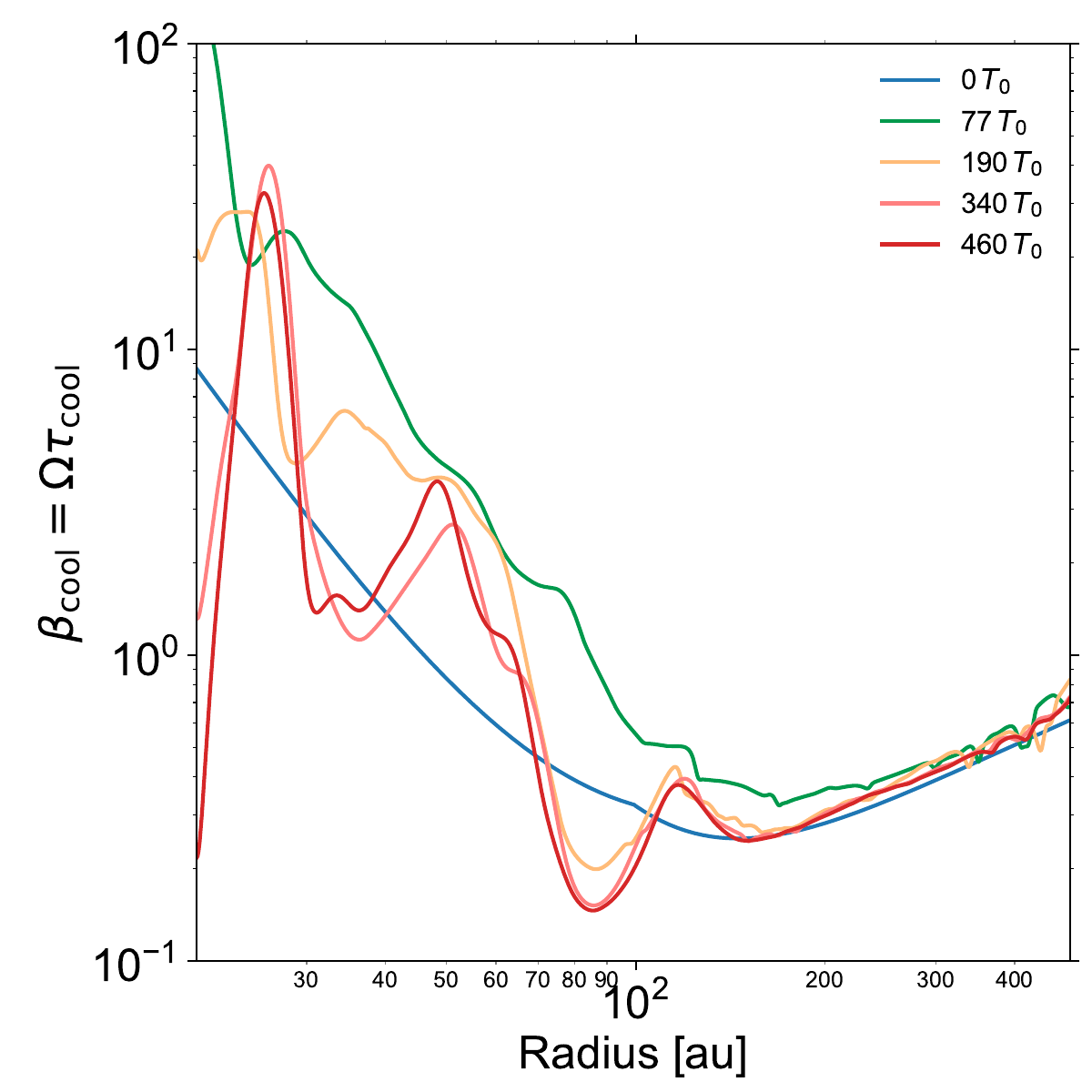}
        \includegraphics[width=0.32\hsize]{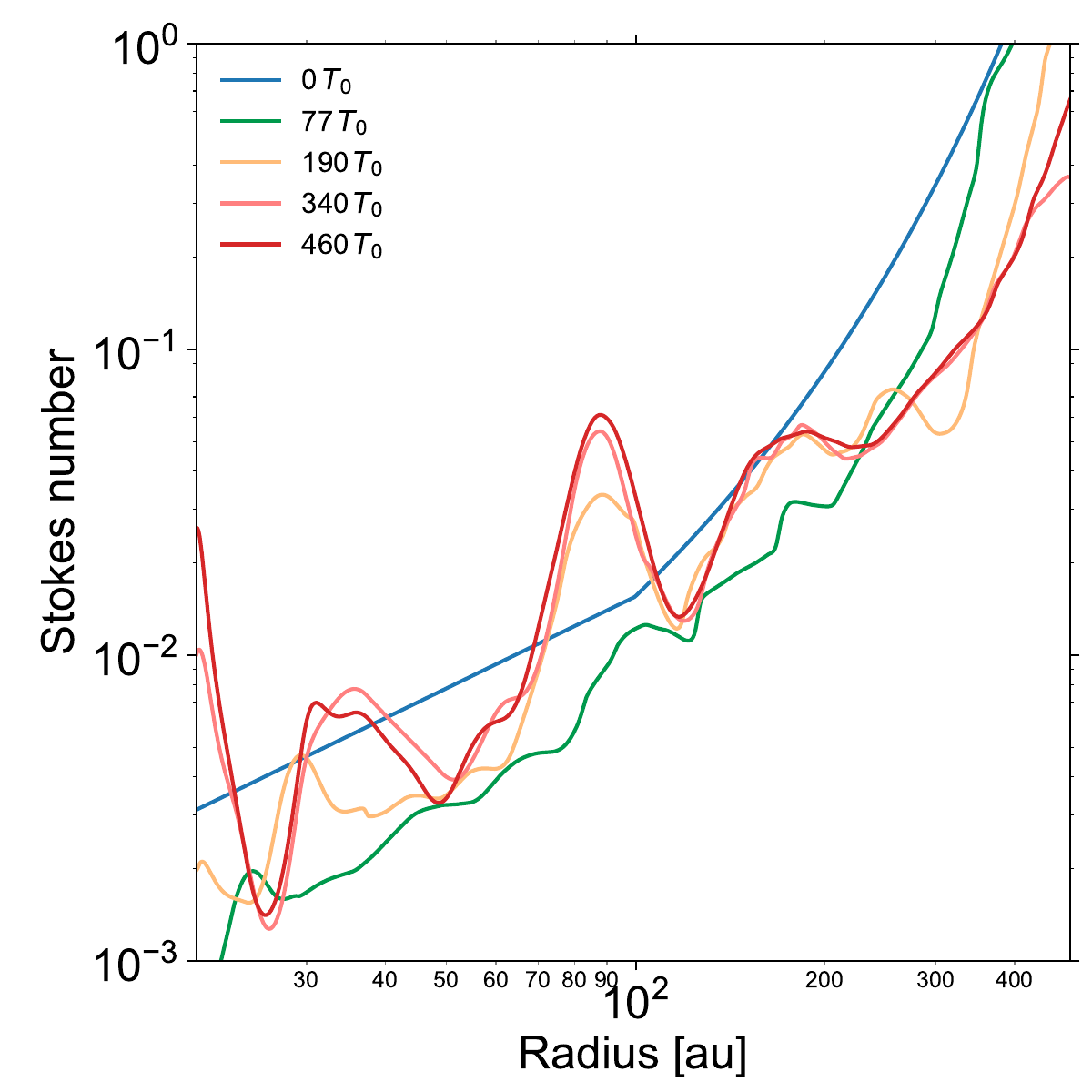}
    \caption{Results of our simulation: all panels except the one in the upper-right corner display azimuthally-averaged radial profiles of disc quantities at the beginning of the simulation and at the four times selected to produce Figures~\ref{fig:fig_2Ddens} and~\ref{fig:fig_RTmedley}-\ref{fig:fig_extra_RT_wiggle} (see labels and legends in the panels). The upper-right panel shows azimuthally-averaged radial profiles of the gas normalised vorticity and of the dust's surface density at 220 $T_0$.
    }
    \label{fig:fig_extra}
\end{figure*}

\noindent{\it Hydrodynamical simulation: 1D radial profiles -- } In \S\ref{subsec:spacetime}, we have chosen to illustrate the global evolution of the disc via space-time diagrams showing how the radial profiles of several disc quantities change over the duration of the simulation. While such space-time diagrams are helpful to display all the information (time, orbital radius, radial profile) in one plot, the precise values of the radial profiles may be difficult to read. We show in Figure~\ref{fig:fig_extra} radial profiles of the gas surface density, temperature, normalised cooling timescale $\beta_{\rm cool}$ (see Eq.~\ref{eq:beta}), dust surface density and Stokes number at the four times selected to produce the panels in Figures~\ref{fig:fig_2Ddens} and~\ref{fig:fig_RTmedley}-\ref{fig:fig_extra_RT_wiggle}. The upper-right panel serves to show the anti-correlation (here at 220 $T_0$) between the normalised vorticity (see \S\ref{subsec:spacetime}) and the dust surface density: minima of the former are maxima of the latter, and vice versa, except perhaps in the disc's outermost region where dust radial drift is fastest.
%FFFFFFFFFFFF
%FFFFFFFFFFFF
\\
\par
\begin{figure*}[!h]
    \centering
        \includegraphics[width=\hsize]{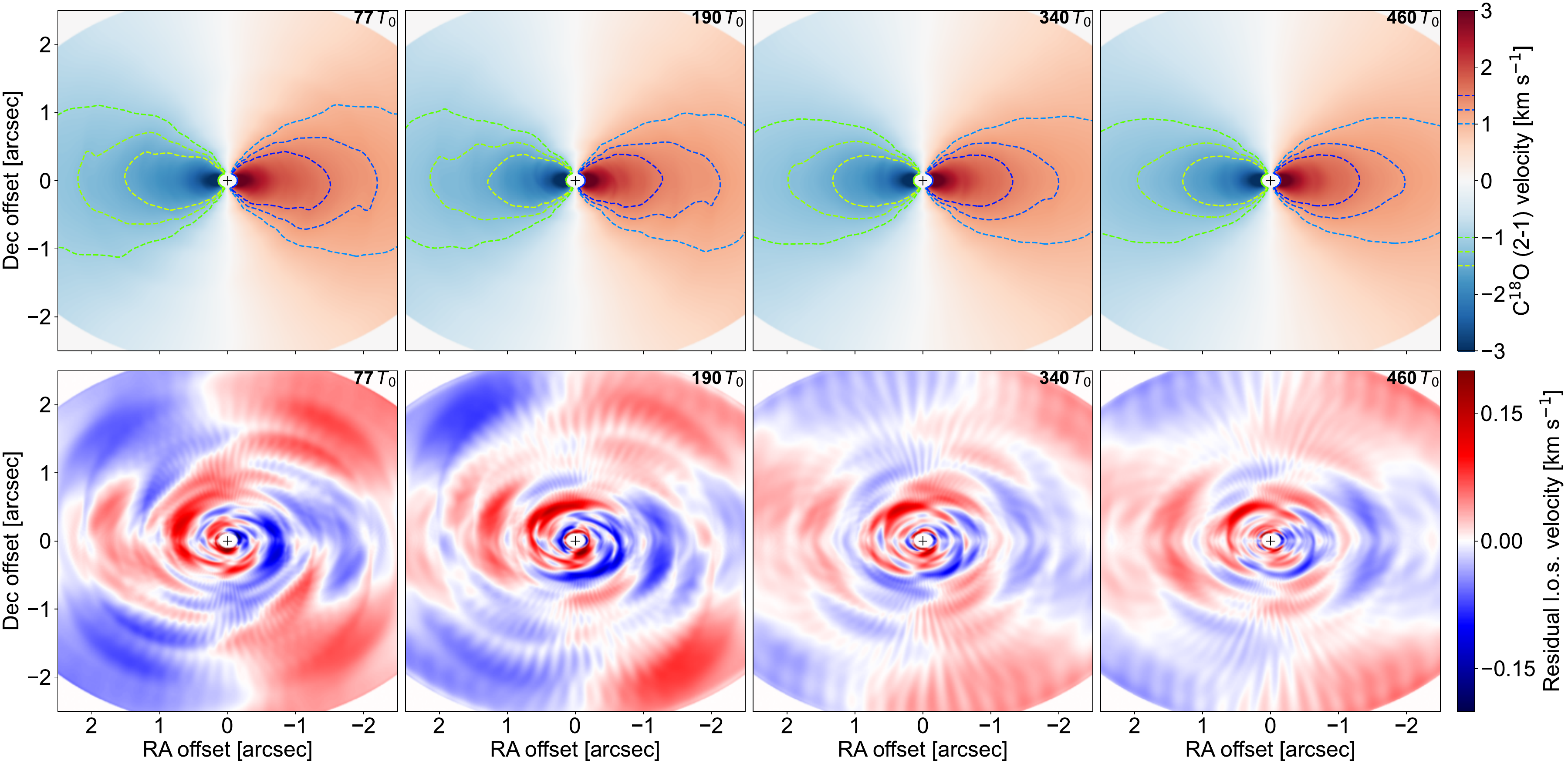}
    \caption{Upper panels: moment 1 maps of the C$^{18}$O (2-1) line emission. Several iso-velocity contours are superimposed as dashed curves, their level is indicated in the colour bar. Lower panels: residual line of sight velocity obtained with the moment 1 maps of the C$^{18}$O (2-1) line emission, from which the projected azimuthal velocity of the gas in the hydrodynamical simulation is subtracted.
    }
    \label{fig:fig_mom1maps}
\end{figure*}

\noindent{\it C$^{18}$O (2-1) line emission: moment 1 maps and residual velocities -- } The upper panels in Figure~\ref{fig:fig_mom1maps} show the moment 1 maps of the C$^{18}$O (2-1) line emission at the four times selected to illustrate the results of our simulation. A few iso-velocity contours have been superimposed to provide a glimpse on how the perturbed velocity structures evolve in time. Spiral structures can also be seen in the central parts of the images. In \S\ref{subsubsec:c18o}, we have explained how we obtained residual velocity maps by subtracting, from the moment 1 maps, a best-fit rotation profile that takes the disc's mass distribution into account. The best-fit was obtained using \href{https://github.com/PlanetFormationLab/eddy}{\texttt{eddy}} \citep{eddy}. For comparison, the lower panels in Figure~\ref{fig:fig_mom1maps} show what we obtain by subtracting instead the projected azimuthal velocity obtained directly from the hydrodynamical simulation. These images should be compared to those in the lower panels of Figure~\ref{fig:fig_RTmedley}. The agreement between both sets of images is very good.
%FFFFFFFFFFFF
%FFFFFFFFFFFF

\end{appendix}

% ================
% REFERENCES
\bibliographystyle{mnras}
\bibliography{reference.bib}
% ================

\end{document}